\documentclass{aa}  

\usepackage{graphicx}
\usepackage{txfonts}

\usepackage{graphicx,graphics,rotating,amssymb,amsmath}
\usepackage{xcolor}
\usepackage{multirow}
\usepackage{longtable}
\usepackage{booktabs}
\usepackage{subfig}
\usepackage{float,capt-of}
\usepackage{natbib}
\usepackage[english]{babel}
\usepackage{morefloats}
\usepackage{color}
\usepackage{sidecap}
\usepackage{epsfig,color}
\usepackage{wrapfig}
 \usepackage{lscape}
\usepackage{url}
\begin{document} 

\newcommand\kms{km~s$^{-1}$}

\title{Enhanced UV radiation and dense clumps in Mrk~231's molecular outflow}

\author{Claudia Cicone \thanks{claudia.cicone@astro.uio.no}
		\inst{1,2}
		\and
		Roberto Maiolino
		\inst{3,4}
		\and
		Susanne Aalto
		\inst{5}
		\and
		Sebastien Muller
		\inst{5}
		\and
		Chiara Feruglio
		\inst{6}
	}

\institute{Institute of Theoretical Astrophysics, University of Oslo, P.O. Box 1029, Blindern, 0315 Oslo, Norway 
	\and INAF-Osservatorio Astronomico di Brera, via Brera 28, 20121, Milan, Italy 
	\and Cavendish Laboratory, University of Cambridge, 19 J. J. Thomson Ave, Cambridge CB3 0HE, UK 
	\and Kavli Institute for Cosmology, University of Cambridge, Madingley Road, Cambridge CB3 0HA, UK 
	\and Department of Space, Earth and Environment, Chalmers University of Technology, Onsala Observatory, SE-439 92 Onsala, Sweden 
	\and INAF - Osservatorio Astronomico di Trieste, via G.B. Tiepolo 11, 34143 Trieste, Italy 
}

\date{Received 27-Sep-2019 / Accepted: 25-Nov-2019 }


\abstract{
We present interferometric observations of the CN(1-0) line emission in Mrk~231 and combine them with previous observations of CO and other H$_2$ gas tracers to study the physical properties of the massive molecular outflow. We find a strong boost of the CN/CO(1-0) line luminosity ratio in the outflow, which is unprecedented compared to any other known Galactic or extragalactic source. For the dense gas phase in the outflow traced by the HCN and CN emissions, we infer $\rm X_{\rm CN}\equiv [CN]/[H_2] > X_{\rm HCN}$ by at least a factor of three, with H$_2$ gas densities of $n_{\rm H_2}\sim10^{5-6}~\rm cm^{-3}$. In addition, for the first time, we resolve narrow spectral features in the HCN(1-0) and HCO$^+$(1-0) high-velocity line wings tracing the dense phase of the outflow. The velocity dispersions of these spectral features, $\sigma_v\sim7-20$~\kms, are consistent with those of massive extragalactic giant molecular clouds detected in nearby starburst nuclei. The H$_2$ gas masses inferred from the HCN data are quite high, $M_{mol}\sim0.3-5\times10^8~M_{\odot}$. Our results suggest that massive, denser molecular gas complexes survive embedded into the more diffuse H$_2$ phase of the outflow, and that the chemistry of such outflowing dense clouds is affected by enhanced UV radiation. 
}


\maketitle


\section{Introduction}\label{sec:introduction}

Mrk~231 is a late-stage merger, classified as ultra-luminous infrared galaxy (ULIRG) ($L_{\rm IR}(\rm 8-1000 \mu m)=1.3 \times 10^{46}~{\rm erg~s^{-1}}$), likely on its path to becoming an elliptical galaxy \citep{Lipari+05}. 
At least $\sim$30\% of the IR luminosity is due to dust heated by an active galactic nucleus (AGN) (\cite{Nardini+10}, but \cite{Veilleux+09a} estimate a higher AGN contribution). Despite its weak emission in the hard X-ray band \citep{Teng+14, Piconcelli+13, Braito+04}, the elevated AGN bolometric luminosity ($L_{\rm AGN} \sim 8 \times 10^{45}~{\rm erg~s^{-1}}$, \citealt{Leighly+14}) makes Mrk~231 the closest quasar known. 
However, most of the IR luminosity traces a young starburst (age ${\rm < 250~Myr}$, \citealt{Davies+07}), forming stars at a rate of ${\rm SFR \gtrsim 200~M_{\odot}~yr^{-1}}$ \citep{Taylor+99}, fed by $\gtrsim10^9$~M$_{\odot}$ of molecular (H$_2$) gas distributed in a kpc-size disk \citep{Downes+Solomon98}. 

The multi-wavelength and multi-scale emission of Mrk~231 is overwhelmingly affected by {\it explosive} phenomena, which are a manifestation of feedback from the star formation and quasar activity, as predicted by theoretical models \citep{Chevalier+Clegg85, Silk+Rees98, Murray+05, Faucher-Giguere+12, Zubovas+King12, Zubovas+King14, Nims+15, Thompson+16, Ishibashi+18} and `observed' in hydrodynamical simulations \citep{DiMatteo+05, Sijacki+07, Costa+15, Costa+18b, Biernacki+Teyssier18, Richings+18a, Schneider+18}.
On nuclear scales, \cite{Feruglio+15} reported observations of a possible X-ray ultra fast wind with $v=20,000$~km~s$^{-1}$. Broad absorption lines (BALs) 
were detected in both low and high ionisation species, classifying Mrk~231 as a FeLoBAL quasar \citep{Lipari+09}. The high-velocity material responsible for the BALs is believed to dominate even the Ly$\alpha$, CIV, and CIII] {\it emission} lines, all showing spectral profiles blueshifted by $-3500$~km~s$^{-1}$ \citep{Veilleux+13}.
Besides the BAL winds, which are confined within $r<10$~pc, Mrk~231 hosts one of the best-studied {\it galaxy-scale} outflows, revealed for the first time in OH absorption by \cite{Fischer+10} and in CO emission by \cite{Feruglio+10}. Using these earlier observations it was estimated that the outflow entrains $\sim3\times10^8$~M$_{\odot}$ of H$_2$ gas and extends out to at least $r\sim1$~kpc into the interstellar medium (ISM) \citep{Cicone+12}.
Blue-shifted NaID \citep{Rupke+Veilleux11,Lipari+09} and H{\sc i}~21cm \citep{Morganti+16,Teng+13} absorption features have also been detected, and they trace the H{\sc i} gas component of the outflow.

At present, most galaxy-scale molecular outflows have been studied exclusively through low-J CO transitions \citep{Cicone+14,Pereira-Santaella+18,Fluetsch+19}.
Such lines, thanks to their low critical density and high abundance of CO in the ISM (provided CO is sufficiently shielded), are ubiquitous tracers of H$_2$ gas, including the diffuse envelopes of molecular clouds that do not host active star formation. On the other hand, transitions from species such as HCN, HCO$^+$, and HNC, because of their higher critical density, should arise mainly from the denser regions of molecular complexes associated with star formation sites \citep{Gao+Solomon04}. Therefore, these molecules (including CN, see below) are commonly referred to as `dense gas' tracers, and we will adopt this terminology throughout the text. However, we note that a significant portion of the emission from such tracers may still be associated with gas that has a density much lower than the nominal critical density ($n_{\rm crit}$) of the corresponding molecular transition. Indeed, $n_{\rm crit}$ is calculated analytically using the optically thin approximation and so it does not account for radiative trapping \citep{Scoville+Solomon74}, which can lower the volume density needed to excite the line by more than one order of magnitude with respect to $n_{\rm crit}$. To deal with line trapping, \cite{Shirley15} propose to use the `effective excitation density' ($n_{\rm eff}$), empirically defined as the gas density that produces a detectable integrated line flux arbitrarily set equal to $\int T_R dv = 1 \rm K~km~s^{-1}$, where $T_R$ is the radiation temperature\footnote{We refer to \cite{Shirley15} for further details on this definition, but we note that there are different nuances of $n_{\rm eff}$ in the literature, see also \cite{Leroy+17}.}. 
Furthermore, the $n_{crit}$ computation in general assumes molecular hydrogen (H$_2$) as the only collisional partner, hence neglecting collisions with electrons and atomic hydrogen. Collisions with electrons may become relevant in a medium where a large fraction of Carbon is ionised but Hydrogen is in the molecular phase, leading to the excitation of molecules such as HCN even in lower density H$_2$ gas ($n_{\rm H_2}\sim10^3~\rm cm^{-3}$, e.g. \cite{Goldsmith+Kauffmann17, Kauffmann+17}).

The first detection of HCN, HCO$^+$, and HNC in the outflow of Mrk~231 was reported by \cite{Aalto+12}. Up to now, this outflow has been studied in eight different species, four detected in absorption by {\it Herschel} (e.g. OH, $^{18}$OH, H$_2$O, OH$^+$, \citealt{Fischer+10,Sturm+11,Gonzalez-Alfonso+14, Gonzalez-Alfonso+18}), and four imaged in emission using (sub-)millimetre interferometry (e.g. CO, HCN, HNC, HCO$^+$, see \citealt{Feruglio+10,Cicone+12,Aalto+12a,Aalto+15,Alatalo15,Feruglio+15,Lindberg+16}).

In this paper we report the detection of exceptionally bright Cyanide Radical (CN) $N=1-0$ emission arising from the outflow of Mrk~231. 
CN is a tracer of high density molecular gas (for $T_k=100\rightarrow10$~K, $n_{\rm crit}\simeq6.4\times10^4\rightarrow4.1\times10^5$~cm$^{-3}$ and $n_{\rm eff}\simeq4.6\times10^3\rightarrow3.8\times10^4$~cm$^{-3}$, assuming a reference column density of $N_{\rm CN}=10^{14}~\rm cm^{-2}$, \cite{Shirley15}). The production of CN is aided by the presence of moderate UV fields, either via the reaction $CH + N \rightarrow CN + H$, since CH is formed rapidly in a medium where both $C^+$ and $H_2$ are abundant \citep{Fuente+93, Meier+15}, or via the photodissociation of HCN: $HCN + \gamma_{UV} \rightarrow CN + H$ \citep{Boger+Sternberg05}. For these reasons, 
CN is expected to be enhanced in photon dominated regions (PDRs), i.e. portions of the clouds where the chemistry is dominated by the interactions with UV photons (e.g. \citealt{Aalto+02, Rodriguez-Franco+98}), but possibly also in the presence of X-rays and/or cosmic rays \citep{Meijerink+07, Garcia-Burillo+10, Harada+15}. 
Observations of Galactic star forming regions have shown that the mere presence of dense molecular gas is not itself a prior for enhanced CN emission. Instead, CN tends to be brighter on the surfaces of molecular clouds exposed to interstellar UV radiation
(PDRs), but also inside molecular clouds embedding young massive stars  \citep{Yamagishi+18}. Notably, CN remains very sensitive to variations in star formation activity even when its emission is averaged across kpc-scale regions in extragalactic sources \citep{Watanabe+14}. 

In this paper we also present the first detection of spectrally-resolved features in the broad wings of the HCN(1-0) and HCO$^+$(1-0) emission lines. Their velocity dispersions are consistent with individual massive molecular clouds entrained in the outflow. We discuss their properties and compare them with known Galactic and extragalactic giant molecular clouds (GMCs).

The paper is organised as follows. The observations are described in $\S$~\ref{sec:obs}. In $\S$~\ref{sec:densetracers} we show the CN(1-0) data and combine them with previous (sub-)millimetre observations of other molecular lines to study the physical properties of the outflowing H$_2$ gas. 
In $\S$~\ref{sec:clumps} we present and discuss new observational evidence for an intrinsic clumpiness of the broad spectral wings of the HCN(1-0) and HCO$^+$(1-0) lines. Our results are discussed in $\S$~\ref{sec:discussion} and summarised in $\S$~\ref{sec:conclusions}.
Throughout the paper we adopt a standard $\Lambda$CDM cosmological model with $H_0$ = 67.8 \kms Mpc$^{-1}$, 
$\Omega_{\Lambda}$ = 0.692, $\Omega_{\rm M}$ = 0.308 \citep{Planck2016}. At the distance of Mrk~231 (redshift $z = 0.04217$, luminosity distance $D_L = 192.4$~Mpc), the physical scale is 0.859~kpc~arcsec$^{-1}$. Uncertainties correspond to $1\sigma$ statistical errors. 

\begin{table*}[tbp]
	\centering \small
	\caption{IRAM PdBI observations of Mrk~231 used in this paper:}
	\label{table:obs}
	\begin{tabular}{lcccccccc}
		\hline
		\hline
		Line(s) & 	$\nu_{\rm rest}$		& $\nu_{\rm obs}$ & Date & Conf. & Clean beam & rms (100~km~s$^{-1}$) & Ref. & Proj. ID \\
				& 	[GHz]		& [GHz]  	  		&   	 & 	 &	[$\arcsec\times\arcsec$]		  &   [mJy~beam$^{-1}$] 			&     		&  \\
		(1)		& 	(2)		& (3)				&	(4)	&	(5)			& (6)		&	(7)								& (8)		& (9)  \\
		\hline
		CN(1-0) &  113.4910$^{\dag}$ &	108.8987$^{\dag}$ &	2010	 & 6C	&	$2.8\times2.6$ &	0.47	& This work & \texttt{UA26} \\
		CO(1-0) & 115.2712  & 	110.6069		&	2009-2010 & 6C/5D 	& $3.2\times2.8$ & 0.25 & \cite{Cicone+12} & \texttt{UA26,T02F} \\
		CO(2-1) & 230.5380	& 	221.2096	 	&	2010 & 6C/5D & 	$1.3\times1.0$		&	0.79	&	\cite{Cicone+12} & \texttt{UB26} \\
		CO(3-2) & 345.7960	& 	331.8038		&	2012 & 6D	& $1.5\times1.3$	&	2.1		&			\cite{Feruglio+15} & \texttt{V087} \\
		HCN(1-0)$^{\ddag}$ 	& 	88.6319			& 85.0455 & 2011 & 6B	& $1.4\times1.2$	&	0.37	&	\cite{Aalto+12} & \texttt{U--D} \\
		HCN(2-1) & 	177.2610	& 	170.0884		&	2012 & 5D	& $2.9\times2.0$	&	0.60		&	\cite{Lindberg+16} &  \texttt{W028}\\
		HCN(3-2) & 	265.8862	& 	255.1275		&	2012-2013 & 6A/6B	& $0.4\times0.3$	&	0.54		&			\cite{Aalto+15} & \texttt{V086,WA85}\\
		\hline
	\end{tabular}
	
	\begin{flushleft}
		{\it Notes:} (1) Molecular transition; (2) rest frequency; (3) observed (sky) frequency; (4) observing dates; (5) configuration of the interferometer; (6) Synthesised beam size achieved after cleaning with natural weighting by using a support encompassing the $3\sigma$ contour levels; (7) 1$\sigma$ rms per beam calculated in channels of $\Delta v=100$~km~s$^{-1}$; (8) relevant references; (9) identification code of the observing programme. $^{\dag}$ Central frequency of the $J=3/2-1/2$ spingroup (i.e. the brightest component of the CN(1-0) doublet); $^{\ddag}$ These observations include also the HCO$^+$(1-0), HNC(1-0), and HC$_3$N(10-9) transitions (see Fig.~\ref{fig:mrk_lines}).
	\end{flushleft}
\end{table*}

\begin{figure*}[tbp]
	\centering
	\includegraphics[clip=true,trim=0cm 0.cm 0cm 0cm,scale=.3,angle=270]{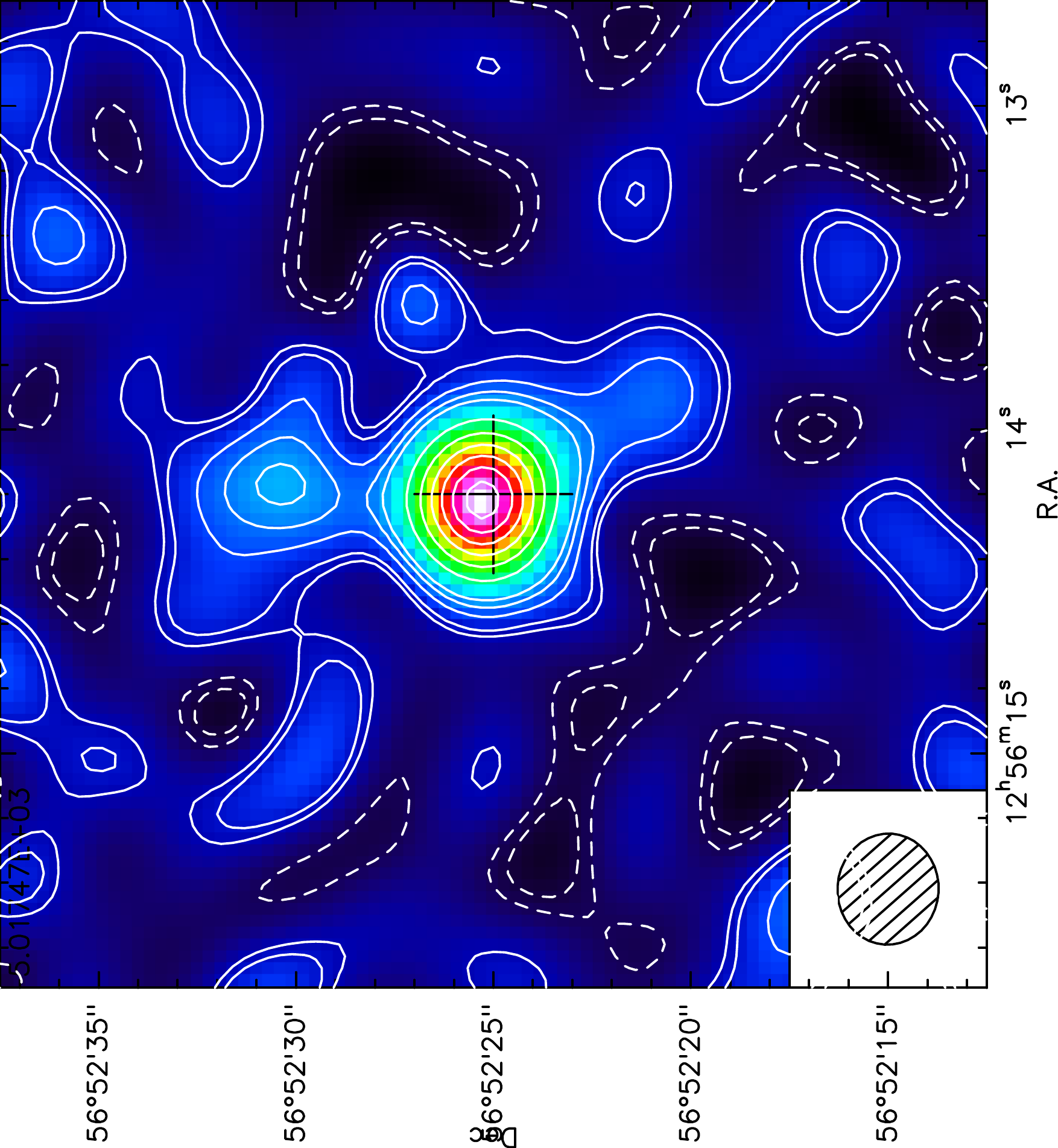}\quad
	\includegraphics[clip=true,trim=0cm 0.cm 0cm 0cm,scale=.3,angle=270]{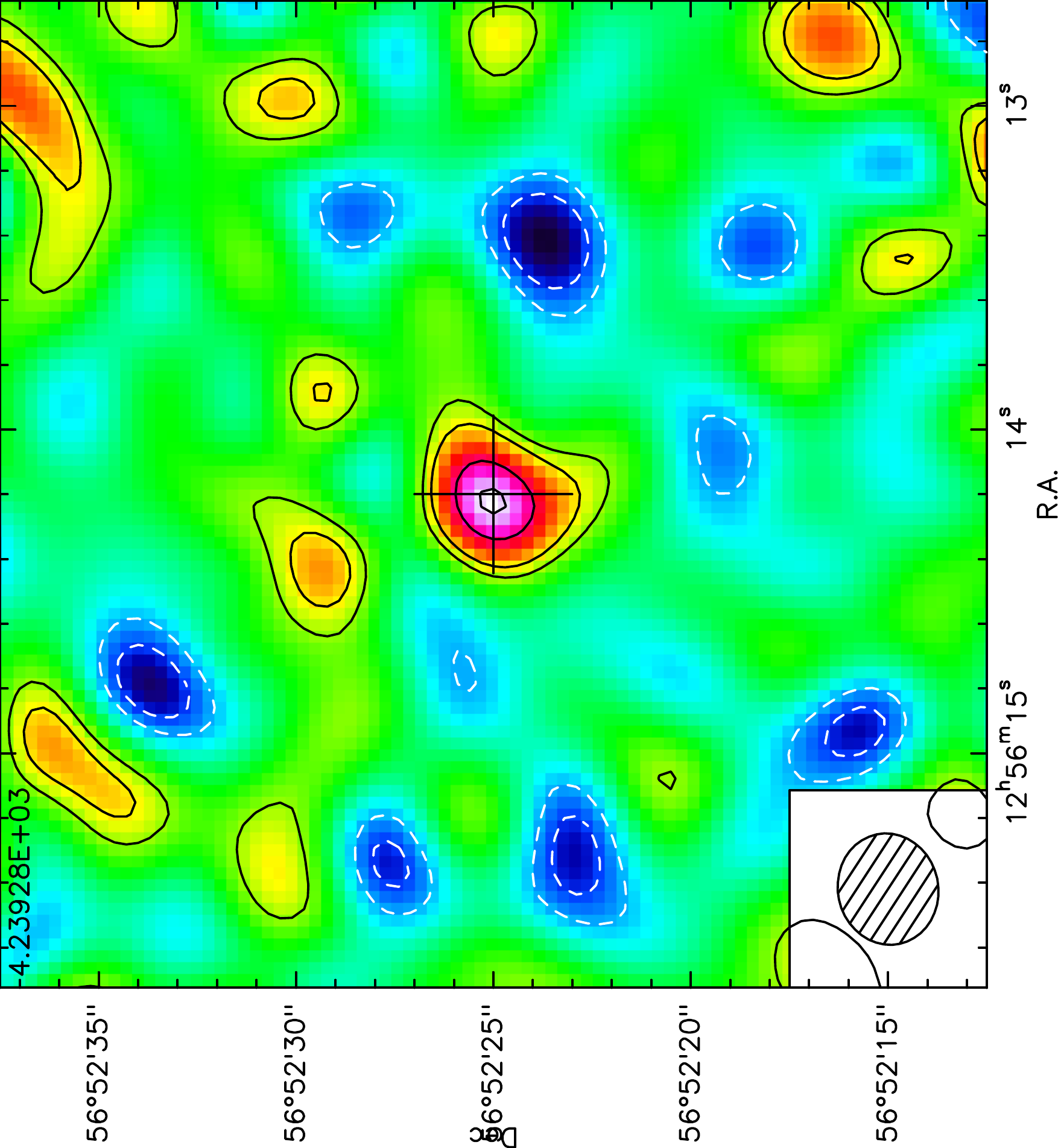}\quad
	\includegraphics[clip=true,trim=0cm 0.cm 0cm 0cm,scale=.3,angle=270]{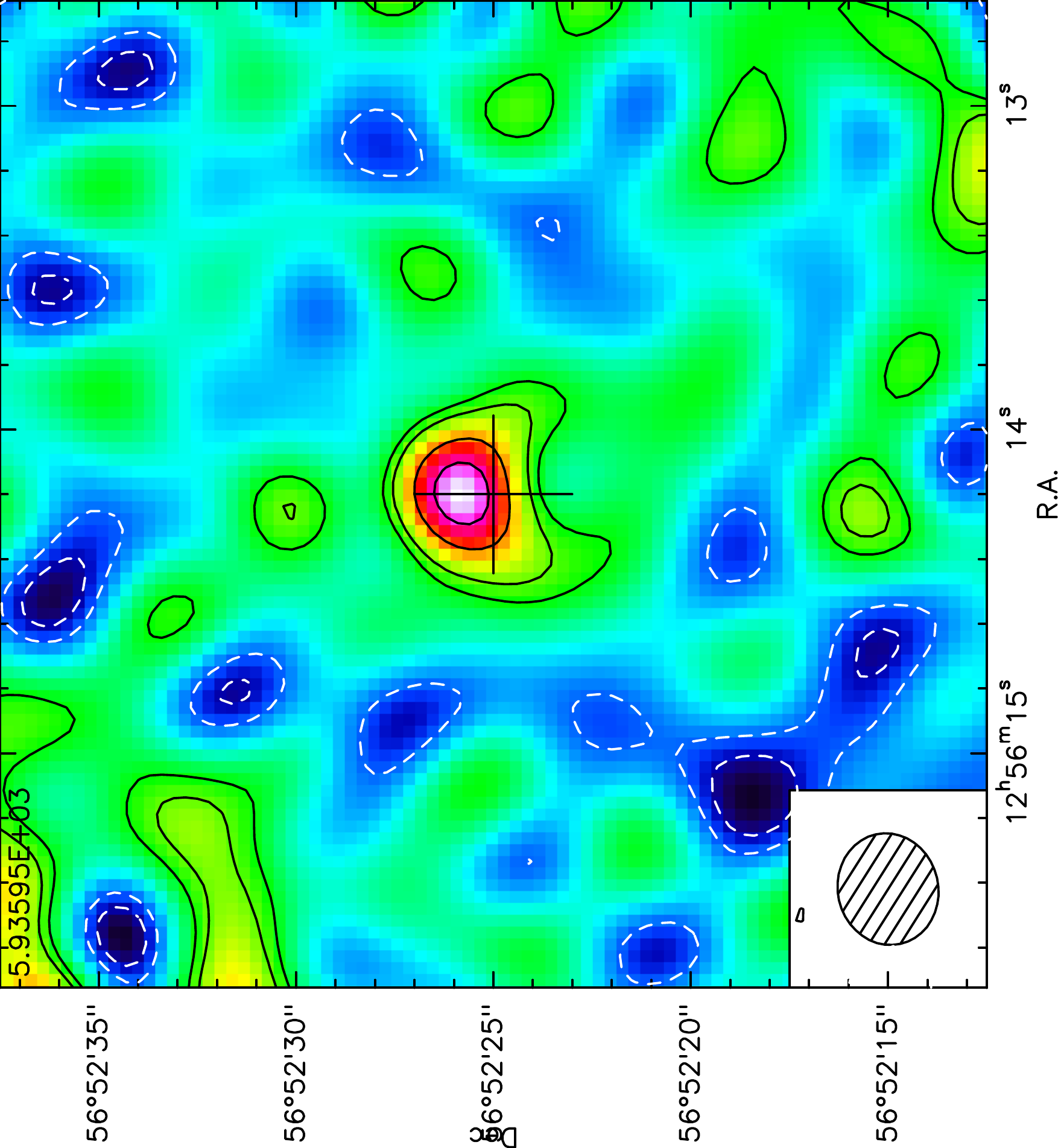}\\
	\caption{IRAM PdBI continuum-subtracted maps of the CN(1-0) line emission. From left to right: (i) total emission integrated between $-800<v~[\rm km~s^{-1}]<1600$, where velocities are calculated with respect to the $J=3/2-1/2$ spingroup (Table~\ref{table:obs}); (ii) blue wing of the $J=3/2-1/2$ line ($-500<v [\rm km~s^{-1}]<-300$); (iii) red wing, integrated between $108.372<\nu_{\rm obs} [\rm GHz] <108.463$, corresponding to velocities $300<v [\rm km~s^{-1}]<550$ with respect to the central frequency of the fainter $J=1/2-1/2$ spingroup (where the red line wing is not affected by blending). All maps are corrected for the primary beam ($45.6\arcsec$ FWHM) and show the central $25''\times25''$ region. The synthesised beam is reported at the bottom-left corner of the maps. Negative contours (dashed lines) correspond to ($-6\sigma$, $-3\sigma$, $-2\sigma$), while positive contours (solid lines) correspond to ($2\sigma$, $3\sigma$, $6\sigma$, $9\sigma$, $n\times12\sigma$), with $n$ =1, 2, etc where $\sigma$ is the rms of the cleaned datacube (see Table~\ref{table:obs}). 
 }\label{fig:cn_maps}
\end{figure*}

\section{Observations}\label{sec:obs}
The $N=1-0$ $J=3/2-1/2$ and $J=1/2-1/2$ spin-doublet transition of CN (hereafter, CN(1-0)) consists of several hyperfine lines (five in the brighter $J=3/2-1/2$ group and four in the $J=1/2-1/2$ group, e.g. \cite{Skatrud+83}), with a frequency separation that is much smaller than the molecular line width of Mrk~231 (maximum separation between the $J=3/2-1/2$ hyperfine levels, $\Delta \nu^{max} = 32.3$~MHz =85~km~s$^{-1}$). The CN(1-0) doublet
was included within the 3.6~GHz-wide spectral bandwidth of the IRAM Plateau de Bure Interferometer (PdBI, now NOrthern Extended Millimeter Array) observations targeting also the CO(1-0) transition and described in \cite{Cicone+12}. However, the CN(1-0) data are shown and analysed for the first time in this paper. 
The phase centre of these observations is RA(J2000) = 12:56:14.200; Dec(J2000)=+56:52:25.0. The other main technical parameters such as rest and observed frequency, date, array configuration, synthesised beam, and rms noise are summarised in Table~\ref{table:obs}.
For data reduction and analysis we used the \texttt{CLIC} and \texttt{MAPPING} packages within the \texttt{GILDAS} software\footnote{\texttt{http://www.iram.fr/IRAMFR/GILDAS}}. The 3mm continuum adjacent to the CO(1-0) and CN(1-0) lines was estimated between $109.3037<\nu_{\rm obs}[\rm GHz]<109.8779$ and $111.2470<\nu_{\rm obs}[\rm GHz]<111.8603$, and subtracted from the $uv$ visibility data prior to imaging and cleaning. 

In this paper we also use archival IRAM PdBI observations of other molecular gas tracers that have been detected in the outflow of Mrk~231, presented in previous publications. Technical details and relevant references for these archival datasets are also reported in Table~\ref{table:obs}.

\section{Results}
\subsection{CN and high density tracers in the outflow}\label{sec:densetracers}

\begin{figure*}[tbp]
	\centering
	\includegraphics[clip=true,trim=4.cm 1.8cm 2.cm 2.5cm,scale=.34,angle=270]{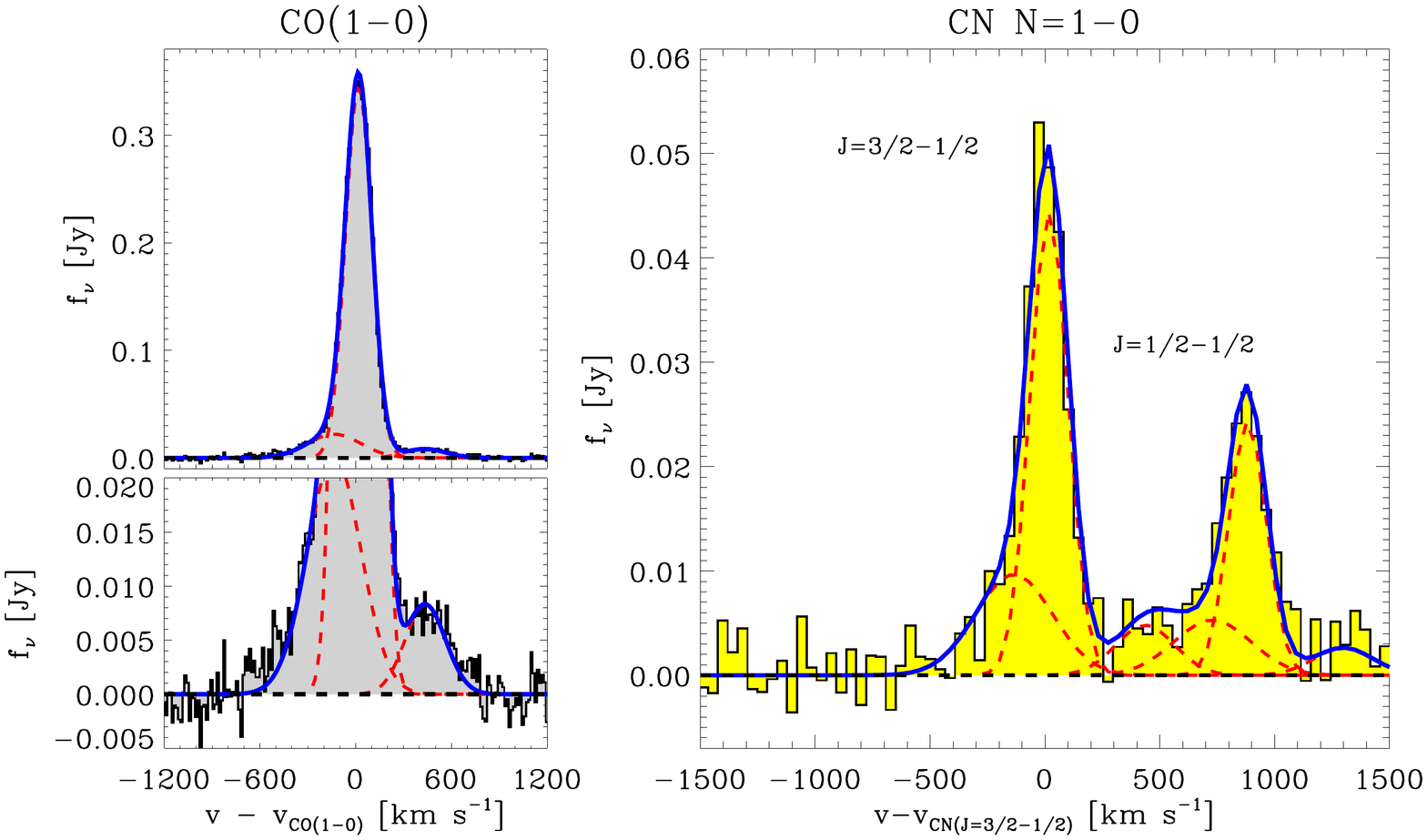}\quad
	\includegraphics[clip=true,trim=4.cm 1.8cm 2.cm 2.5cm,scale=.34,angle=270]{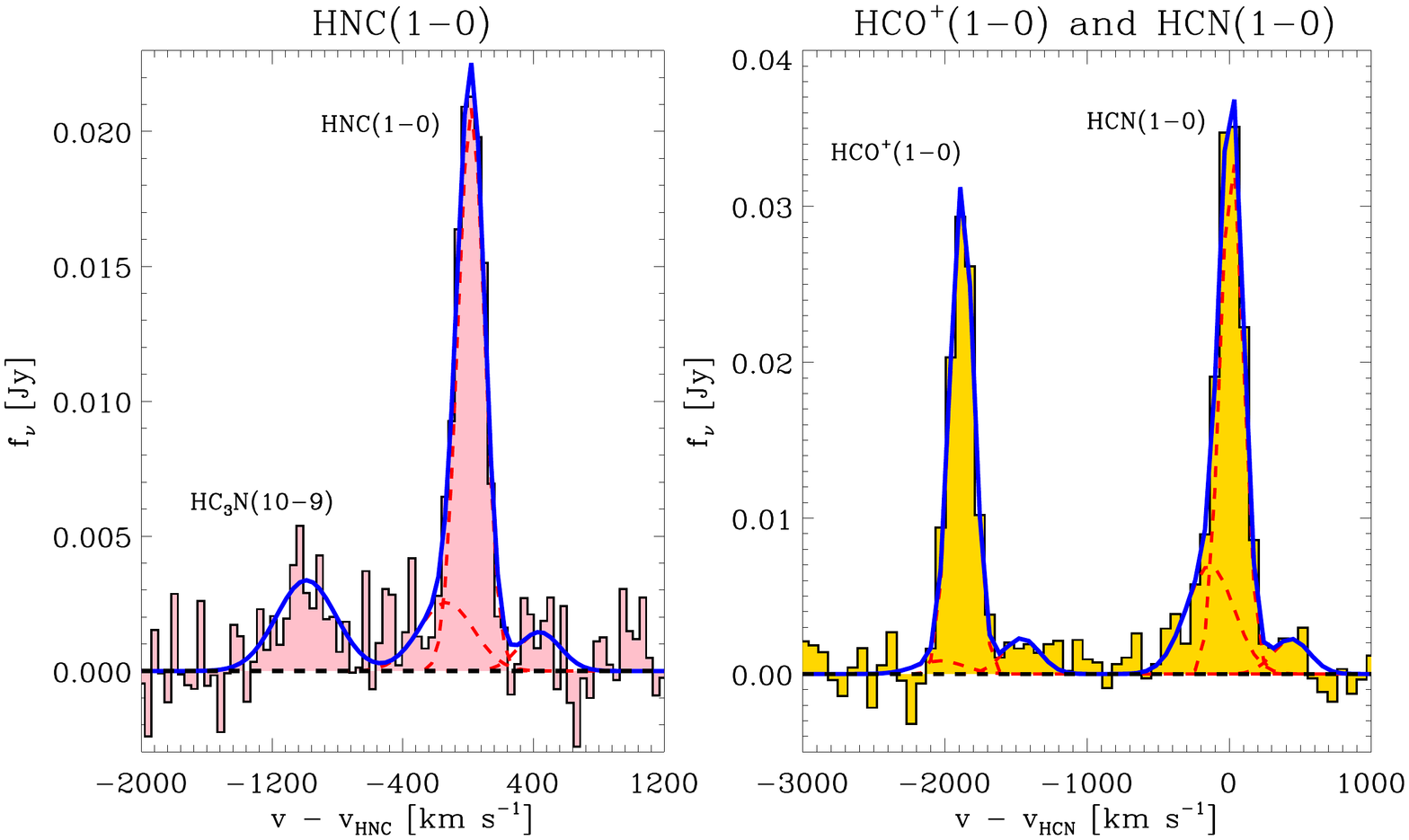}\\
	\includegraphics[clip=true,trim=4.cm 1.8cm 2.cm 2.5cm,scale=.34,angle=270]{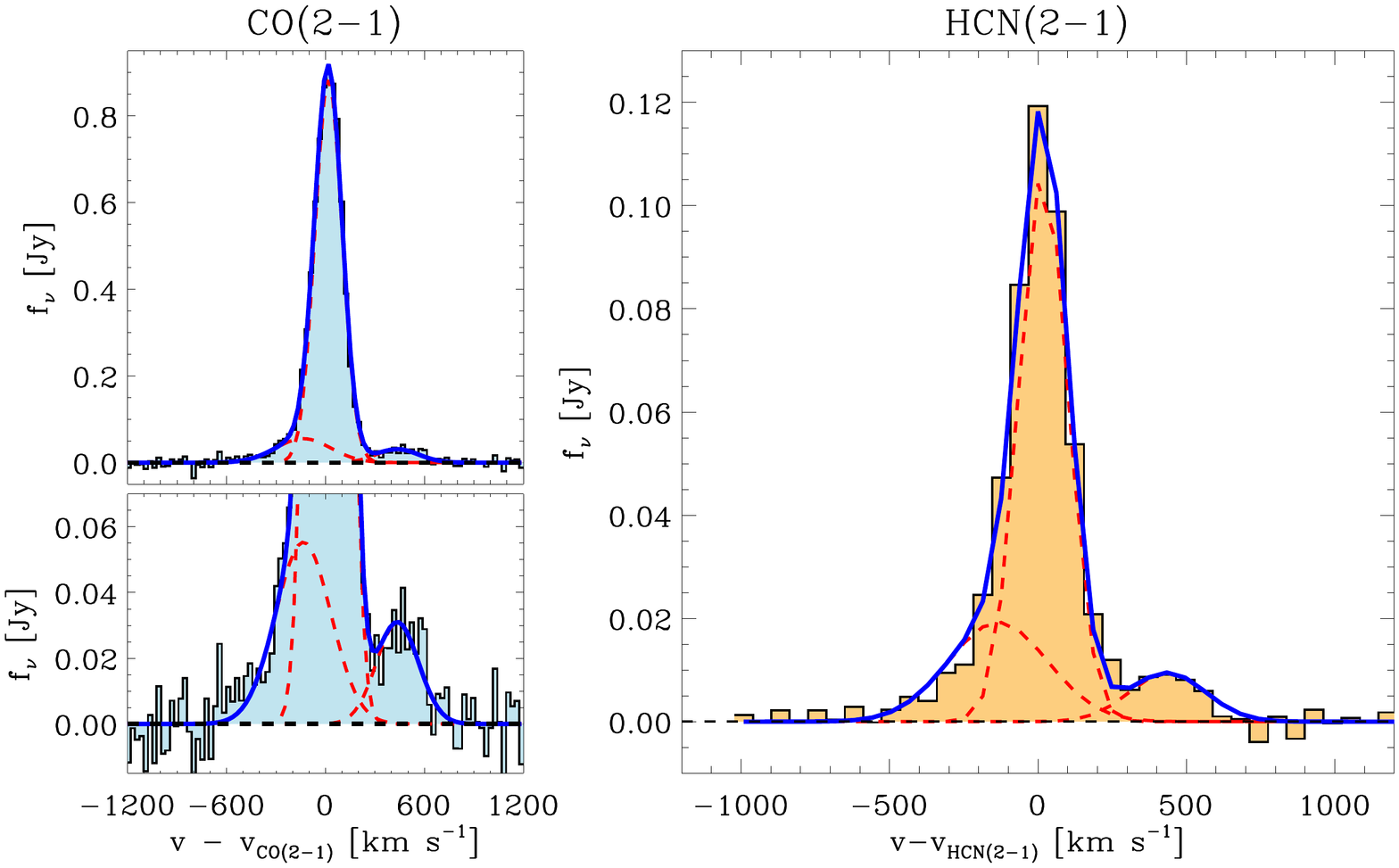}\quad
	\includegraphics[clip=true,trim=4.cm 1.8cm 2.cm 2.5cm,scale=.34,angle=270]{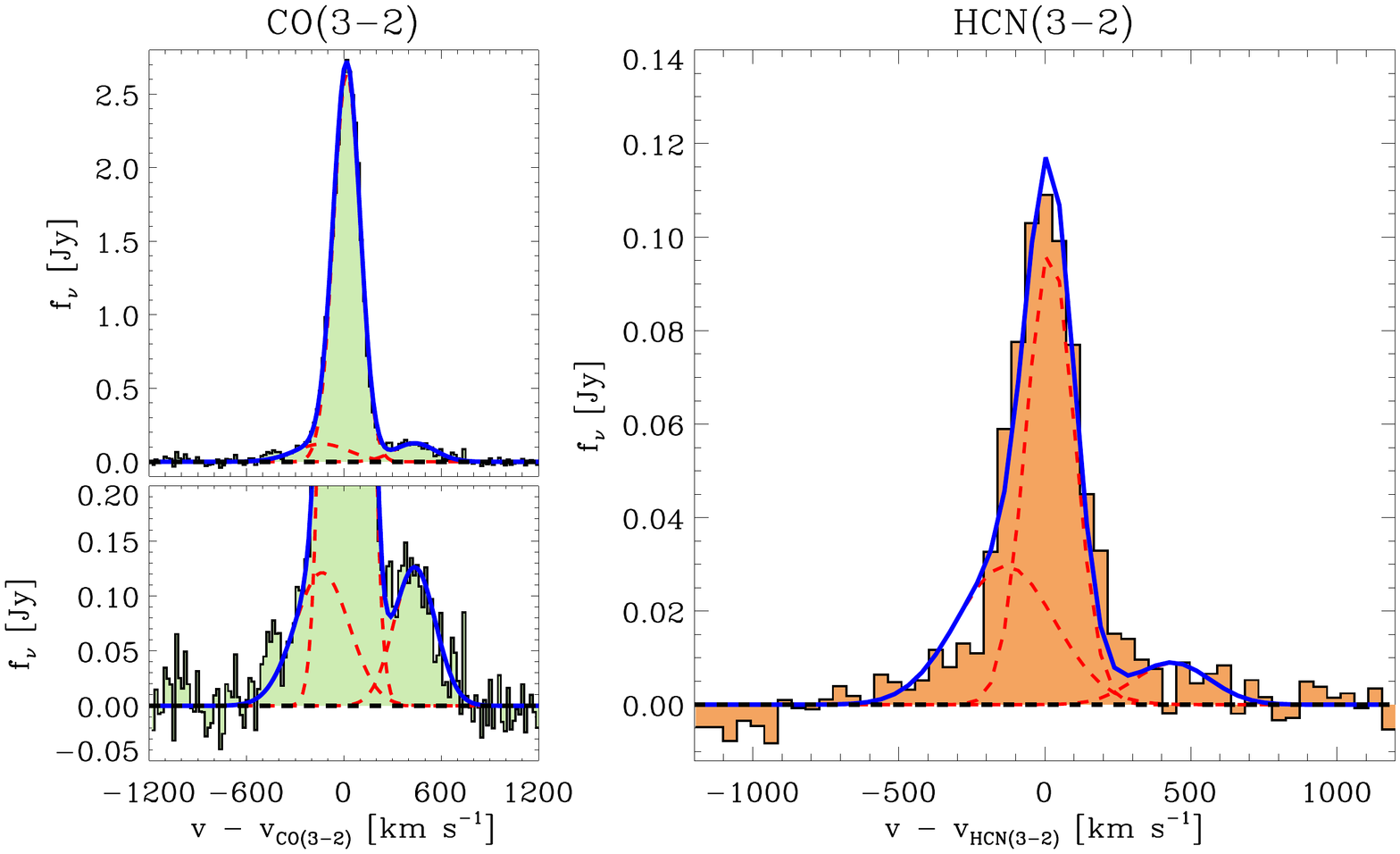}\\
	\caption{Total spectra - and simultaneous fit - of the CO(1-0), CN(1-0), HNC(1-0), HCO$^+$(1-0), HCN(1-0), CO(2-1), HCN(2-1), CO(3-2), and HCN(3-2) molecular lines in Mrk~231. Overlaid on the data, we show the individual Gaussians employed in the simultaneous fit (up to three components for each line, red dashed curves) and the total best-fit curve (blue solid lines). Only the fainter HC$_3$N(10-9) line was fitted by a single Gaussian with unconstrained velocity and width. The rms values per spectral channel are: 1.7~mJy per $\delta v=14$~\kms (CO(1-0));  2.3~mJy per $\delta v=43$~\kms (CN(1-0)); 1.7~mJy per $\delta v=40$~\kms (HNC(1-0)); 1.4~mJy per $\delta v=70$~\kms (HCO$^+$ and HCN(1-0)); 8~mJy per $\delta v=26$~\kms (CO(2-1)); 
		1.2~mJy per $\delta v=60$~\kms (HCN(2-1)); 26~mJy per $\delta v=18$~\kms (CO(3-2)); 5~mJy per $\delta v=47$~\kms (HCN(3-2)).}\label{fig:mrk_lines}
\end{figure*}

\begin{table*}[tbp]
	\centering
	\caption{Line fluxes resulting from the simultaneous fit to the total integrated spectra:}
	\label{table:fit_total}
	\begin{tabular}{lcccc}
		\hline
		\hline
		Transition 		&	Narrow core$^{(*)}$  	&  Blue Wing$^{(\dag)}$ &	Red Wing$^{(\ddag)}$	& 		Total line	\\
		&	[Jy~\kms] 						&	    [Jy~\kms] 		    	&   		 [Jy~\kms] 		&		 [Jy~\kms] 	\\
		\hline
		CO(1-0)					&	$71.0\pm0.6$	& $9.2\pm0.7$	&    $2.6\pm0.2$ &	$83.0\pm1.0$\\ 
		CN(1-0)  $J=3/2-1/2$ &	$9.2\pm0.4$	&  $4.1\pm0.5$	  &    	$1.5\pm0.3$			&	$14.8\pm0.8$ \\ 
		CN(1-0)  $J=1/2-1/2$  &	$5.0\pm0.3$ & 	$2.2\pm0.3$	&    $0.8\pm0.2$			& $8.1\pm0.5$	\\
		HNC(1-0) 		&	$4.3\pm0.3$		&	$1.1\pm0.4$	&    $0.5\pm0.2$		&	$5.8\pm0.5$	\\ 
		HCO$^+$(1-0)	&	$6.4\pm0.3$	  &	$0.4\pm0.4$	&    $0.7\pm0.3$ 	&	$7.5\pm0.6$		\\ 
		HCN(1-0)		&	$6.9\pm0.3$		& $2.9\pm0.4$	&    $0.7\pm0.3$	& $10.6\pm0.6$	\\ 
		CO(2-1)			&	$182.0\pm1.7$	&	$23\pm2$	&    $9.9\pm1.1$	&	$215\pm3$	\\ 
		HCN(2-1)		&	$22.0\pm0.5$	&	$8.1\pm0.6$	&    $3.0\pm0.3$	&	$33.2\pm0.8$ \\ 
		CO(3-2)			&	$545\pm4$	&	$51\pm5$	&    $40\pm4$	&	$637\pm8$	\\ 
		HCN(3-2)		&	$20.1\pm1.0$	&	$12.5\pm1.4$	&    $2.9\pm0.8$	&	$35.4\pm1.9$ \\ 
		\hline
		HC$_3$N(10-9)     & 						&							&						& $1.5\pm0.5$ 						\\
		\hline
	\end{tabular}
	
	\begin{flushleft}
		{\bf Notes:} 
		$^{(*)}$ Narrow core best-fit kinematics: $v=18.0\pm0.2$~\kms; $\sigma_v=82.3\pm0.3$~\kms. 
		$^{(\dag)}$ Blue wing best-fit kinematics: $v= -135\pm16$~\kms ; $\sigma_v=168\pm8$~\kms.   
		$^{(\ddag)}$ Red wing best-fit kinematics: $v=437\pm7$~\kms ; $\sigma_v= 128\pm7$~\kms.  
		The HC$_3$N(10-9) line is modelled by a single Gaussian, with unconstrained width and velocity, shifted by $v=-990\pm50$~\kms with respect to the HNC(1-0) line, with 
		$\sigma_v = 180\pm50$~\kms and $S^{peak}=3.4\pm0.8$~mJy. 
	\end{flushleft}
\end{table*}

The continuum-subtracted maps of the total CN(1-0) line emission and of the blue- and red-shifted CN(1-0) components are presented in Figure~\ref{fig:cn_maps}. For the line wings, we selected velocities that are not affected by blending of the two CN(1-0) spingroups. Besides the well-known central concentration of H$_2$ gas embedding the molecular disk already studied by \cite{Downes+Solomon98}, we identify a northern component, detected in CN at the 12$\sigma$ level, which is spatially extended and offset by $\sim3-7$~kpc with respect to the nucleus of Mrk~231 (left panel of Fig.~\ref{fig:cn_maps}). To our knowledge, this is the first detection of such an extended molecular gas structure outside the main disk of Mrk~231. In Appendix~\ref{sec:appendixA} we show the CN(1-0) and CO(1-0) spectra extracted from a region encompassing the northern CN feature. The properties of this structure, i.e. its spectral line profile, spatial extent, and CN/CO(1-0) line luminosity ratio, suggest that it may represent an $>5$~kpc-size extension of the molecular outflow. Further supporting such interpretation, we note that previous CO and HCN observations have already evidenced blue-shifted and red-shifted features extending by $\sim2$~kpc north of the nucleus of Mrk~231 \citep{Cicone+12, Aalto+12, Feruglio+15}, i.e. in a region lying between the newly-discovered northern CN structure and the well-known central kpc component of the outflow.

The total CN(1-0) spectrum of Mrk~231 is presented in Fig~\ref{fig:mrk_lines}, together with the CO(1-0), HNC(1-0), HCO$^+$(1-0), HCN(1-0), CO(2-1), HCN(2-1), CO(3-2), and HCN(3-2) line spectra. The sizes of the emitting regions differ slightly among different transitions, due to a combination of (i) different beam sizes, since higher resolution data (e.g. taken with configurations A and B of the IRAM PdBI, see Table~\ref{table:obs}) have lower sensitivity to large-scale structures because of interferometric filtering, and (ii) physical conditions, where higher excitation lines tend to be more compact than CO(1-0), as already noted for the outflow of Mrk~231 by \cite{Cicone+12}. For this reason, we employed slightly different aperture sizes for different tracers, chosen to maximise the total emission detected in the corresponding primary beam-corrected datacubes. More specifically, the spectra shown in Fig~\ref{fig:mrk_lines} were extracted from circular apertures with diameters of 10$\arcsec$ for CO(1-0), CN(1-0), CO(2-1), and HCN(2-1); 5$\arcsec$ for CO(3-2); 4$\arcsec$ for HCN(1-0), HCO$^+$(1-0), and HNC(1-0); and 3$\arcsec$ for HCN(3-2). We note however that adopting the same 10$\arcsec$-size aperture for all lines would not change our results, but it would just result in slightly noisier HCN(1-0), HNC(1-0), HCO$^+$(1-0), and HCN(3-2) spectra.

The striking similarity between the spectral profiles of all these different molecular transitions allows us to perform a simultaneous fit, whose results are overlaid on the data in Fig~\ref{fig:mrk_lines}. The fit employs three Gaussian components, which are tied to have the same velocity and width for each line. Such simultaneous fitting procedure allows us to robustly deblend the broad wings of the two CN spin-groups. The amplitude ratio between the $J=1/2-1/2$ and $J=3/2-1/2$ components is fixed for each Gaussian, and constrained by the fit to be $0.55\pm0.03$, which is close to the theoretical optically thin value ($0.5$), as already noted by \cite{Henkel+14} using lower S/N single-dish CN(1-0) observations of Mrk~231. All other best-fit values are listed in Table~\ref{table:fit_total}.  

\begin{figure}[tbp]
	\centering
	\includegraphics[clip=true,trim=4.2cm 2.8cm 3cm 8.cm,scale=.45,angle=270]{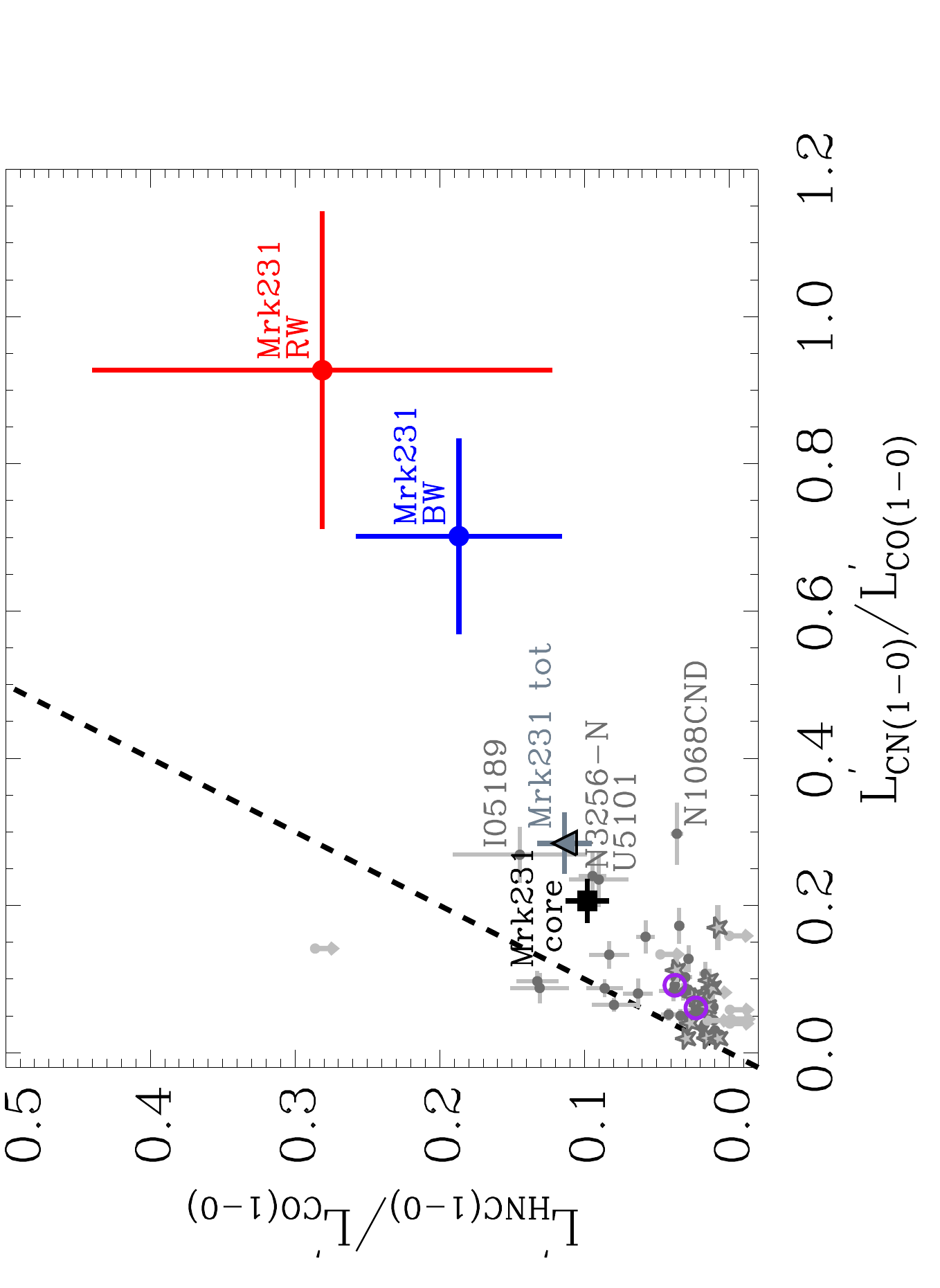}\quad
	\includegraphics[clip=true,trim=4.2cm 3cm 3cm 8.cm,scale=.45,angle=270]{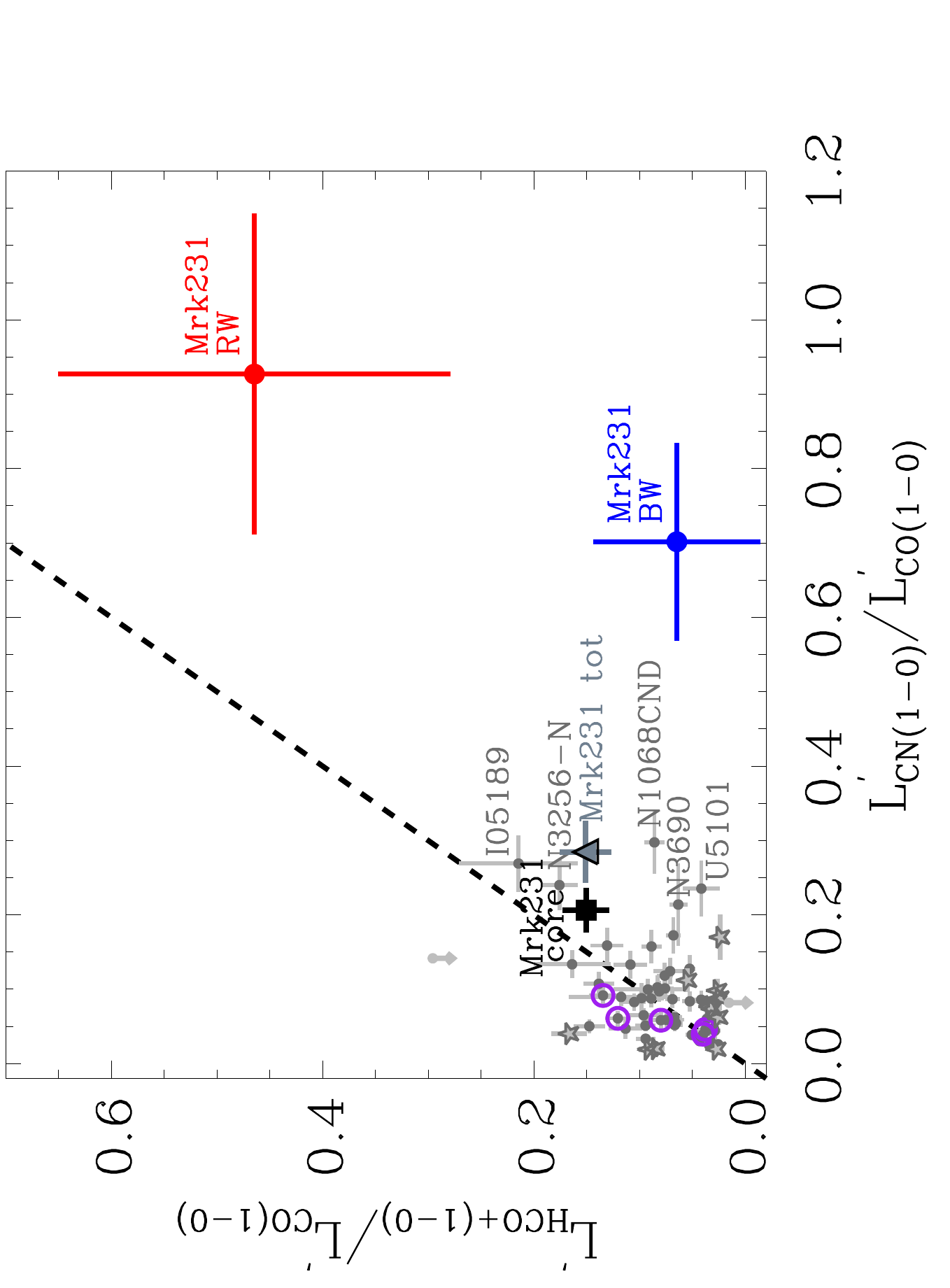}\quad
	\includegraphics[clip=true,trim=4.2cm 3cm 3cm 8.cm,scale=.45,angle=270]{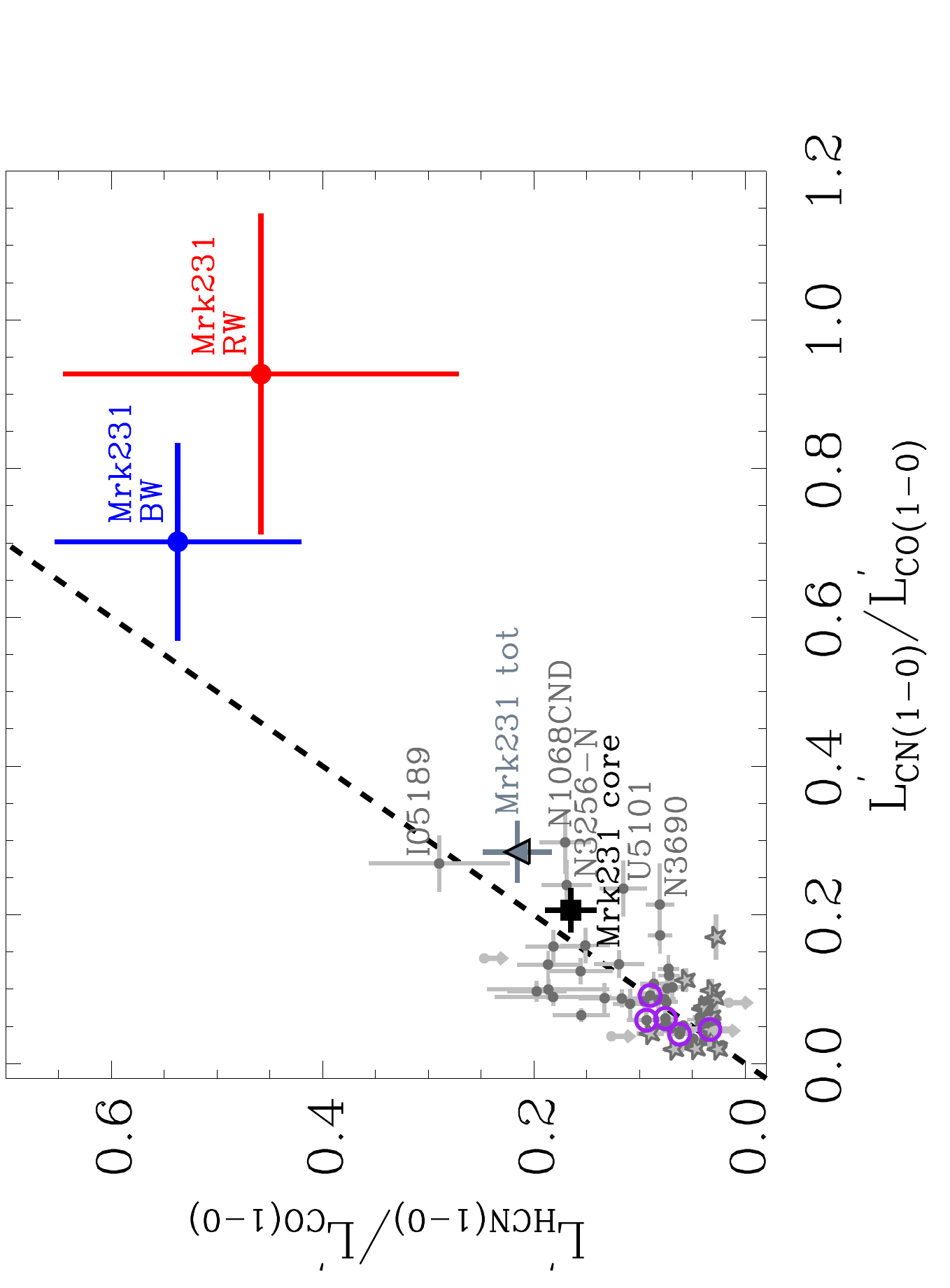}\\
	\caption{CN/CO(1-0) versus HNC/CO(1-0) ({\it top}), HCO$^+$/CO(1-0) ({\it middle}), and HCN/CO(1-0) ({\it bottom}) line luminosity ratios for the different line components of Mrk~231 (narrow core, blue wing, red wing, and total emission),
	including literature samples of nearby galaxies (grey circles) and GMCs (grey stars). Measurements corresponding to extragalactic molecular outflows (NGC~253 and NGC~3256, see Table~\ref{table:literature}) are highlighted with purple circles. The CN(1-0) luminosity corresponds to the sum of the two spingroup components. When only the brightest $J=3/2-1/2$ spingroup line flux is available in the literature, we extrapolated the total CN(1-0) flux by multiplying this value by 1.5. The dashed line indicates the 1:1 relation. All errorbars include an 10\% calibration uncertainty, which was added in quadrature to the statistical error (1$\sigma$). Upper limits are shown at the 3$\sigma$ level. Sources with $L^{\prime}_{\rm CN(1-0)}/L^{\prime}_{\rm CO(1-0)}>0.2$ are labelled in the plot.}\label{fig:line_ratio_plots}
\end{figure}

\begin{figure}[tb]
	\centering
	\includegraphics[clip=true,trim=5cm 3cm 3cm 8.cm,scale=.55,angle=270]{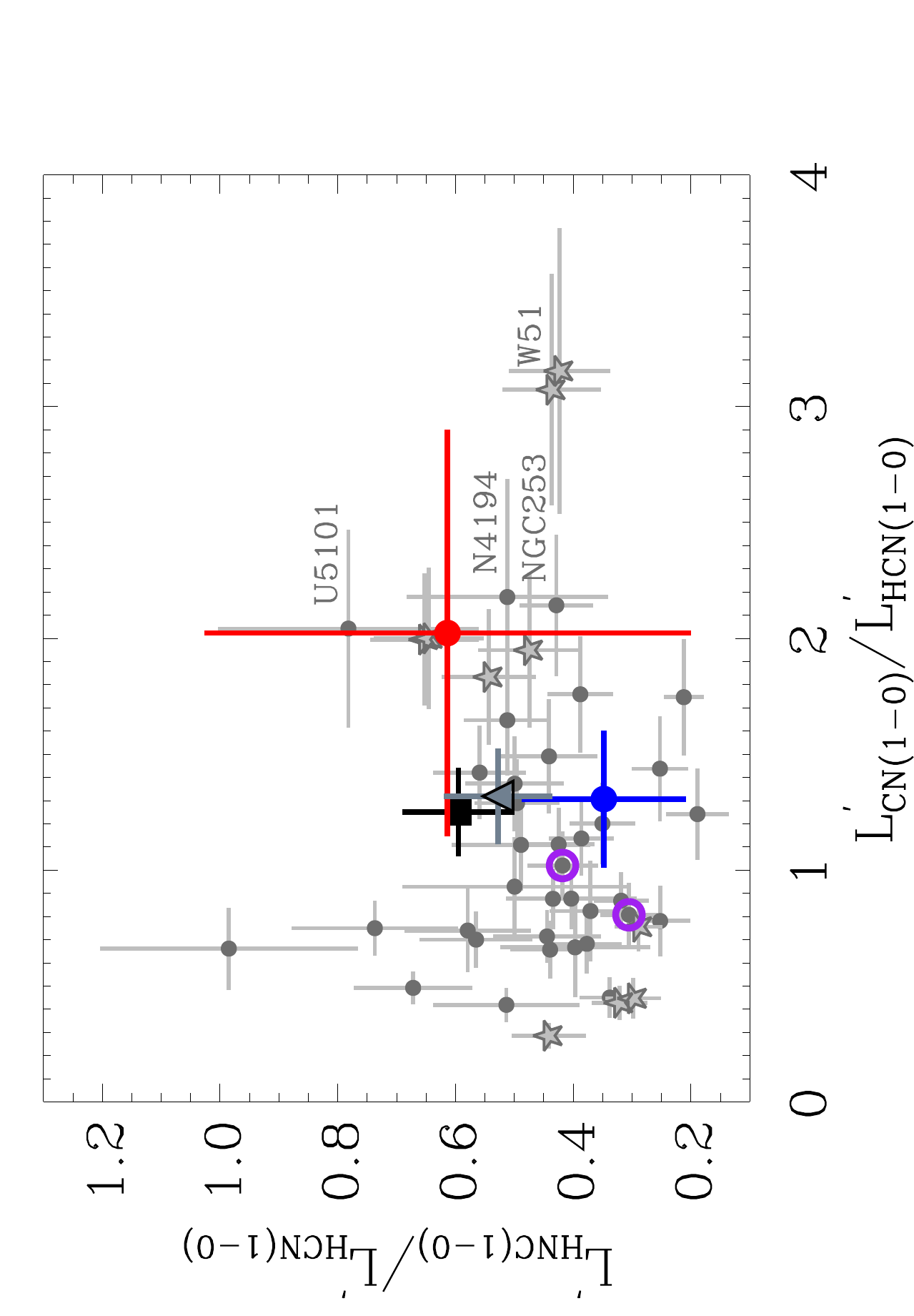}\\
	\caption{CN/HCN(1-0) versus HNC/HCN(1-0) line luminosity ratios for the different line components of Mrk~231, compared to a sample of local galaxies and Galactic GMCs from the literature. See also caption of Fig.~\ref{fig:line_ratio_plots}.}\label{fig:line_ratio_plots_2}
\end{figure}

Figure~\ref{fig:mrk_lines} shows that the broad wings are significantly more prominent in the dense gas tracers (e.g. CN, HCN) than in (low-J) CO transitions. For example, the CN(1-0) and HCN(1-0) wings-to-core flux ratios are higher respectively by a factor of four and three compared to CO(1-0). 

The enhancement of high density tracers in Mrk~231's outflow is further explored in Figure~\ref{fig:line_ratio_plots}, where we plotted, separately for the three Gaussian components fitted to Mrk~231's lines, the CN/CO ($J=1-0$) against the HNC/CO, HCO$^+$/CO, and HCN/CO ($J=1-0$) line luminosity ratios (i.e. the ratios between the line luminosities $L^{\prime}_{\rm line}$ in units of [K~km~s$^{-1}$~pc$^2$], calculated using the definition by \cite{Solomon+05}, which are equivalent to brightness temperature ratios). For comparison, Figure~\ref{fig:line_ratio_plots} includes also a sample of nearby galaxies as well as Galactic and extragalactic GMCs drawn from the literature (all literature values are reported in Table~\ref{table:literature} in Appendix~\ref{sec:AppendixB}).
Most of the literature data points correspond to integrated line measurements, without differentiating between disk and outflow components.
The only exceptions are the galaxies NGC~253 and NGC~3256, for which molecular line fluxes have been measured at different locations by \cite{Meier+15} and \cite{Harada+18}, including a few positions coincident with their molecular outflows (highlighted with purple circles in Fig~\ref{fig:line_ratio_plots}). 

It is evident from Fig.~\ref{fig:line_ratio_plots} that $L^{\prime}_{\rm CN(1-0)}/L^{\prime}_{\rm CO(1-0)}$ is significantly boosted in both the blue- and red-shifted sides of Mrk~231's outflow, compared to the value measured in 
the narrow core of the lines, which should trace mainly the molecular disk.
An enhancement with respect to the core is measured also in $L^{\prime}_{\rm HNC(1-0)}/L^{\prime}_{\rm CO(1-0)}$ and 
$L^{\prime}_{\rm HCN(1-0)}/L^{\prime}_{\rm CO(1-0)}$ for both line wings, and in
$L^{\prime}_{\rm HCO^{+}(1-0)}/L^{\prime}_{\rm CO(1-0)}$ for the red wing (but not for the blue wing). However, among all line ratios explored here, $L^{\prime}_{\rm CN(1-0)}/L^{\prime}_{\rm CO(1-0)}$ shows the largest enhancement in the line wings relative to the core.

Furthermore, the $L^{\prime}_{\rm CN(1-0)}/L^{\prime}_{\rm CO(1-0)}$ values of $0.70\pm0.13$ and $0.9\pm0.2$ measured respectively in the blue and red wings are significantly higher than the corresponding $L^{\prime}_{\rm HNC(1-0)}/L^{\prime}_{\rm CO(1-0)}$ ($0.19\pm0.07$ and $0.28\pm0.16$ for the blue and red wings), $L^{\prime}_{\rm HCO+(1-0)}/L^{\prime}_{\rm CO(1-0)}$ ($0.46\pm0.19$ for the red wing) and $L^{\prime}_{\rm HCN(1-0)}/L^{\prime}_{\rm CO(1-0)}$ ($0.54\pm0.12$ and $0.46\pm0.19$ for the blue and red wings) values. Hence, CN shows the highest luminosity advantage in the line wings with respect to CO. Quite strikingly, the red wing of Mrk~231 is as bright in CN(1-0) as in CO(1-0) emission. Such high CN/CO line luminosity ratios are unprecedented if compared to any other extragalactic or Galactic measurement available in the literature. In particular, neither NGC~253 nor NGC~3256 show an enhancement of CN (or other dense gas tracers) in the outflow comparable to that seen in Mrk~231, suggesting that the physical state of the molecular gas in outflow
can differ significantly from source to source. The CN(1-0) line emission is exceptionally bright in the outflow of Mrk~231, not only compared to Mrk231's own molecular disk (traced by the `core' component shown in Fig~\ref{fig:line_ratio_plots}), but also compared to any other known astrophysical environment. 

However, while Mrk~231's outflow shows an unprecedented enhancement of high density tracers such as CN(1-0) compared to the CO(1-0) line emission (which should be a good proxy of the total H$_2$ gas mass), the line luminosity ratios computed between different dense gas tracers (e.g. CN/HCN, HNC/HCN) are not uncommon compared to other known astrophysical sources. Indeed, Figure~\ref{fig:line_ratio_plots_2} shows that the $L^{\prime}_{\rm CN(1-0)}/L^{\prime}_{\rm HCN(1-0)}$ and $L^{\prime}_{\rm HNC(1-0)}/L^{\prime}_{\rm HCN(1-0)}$ values measured in the wings and in the core of Mrk~231's lines are formally consistent with each other (although $L^{\prime}_{\rm CN(1-0)}/L^{\prime}_{\rm HCN(1-0)}$ is slightly higher in the red wing than in the other components). Furthermore, both line luminosity ratios fall 
within the range of values measured in the literature sample.

\subsection{Spectrally-resolved clumps in the HCN and HCO$^+$ wings}\label{sec:clumps}

\begin{figure*}[tbp]
	\centering
	\includegraphics[clip=true,trim=3.cm 1.6cm 7cm 2.cm,scale=.68,angle=270]{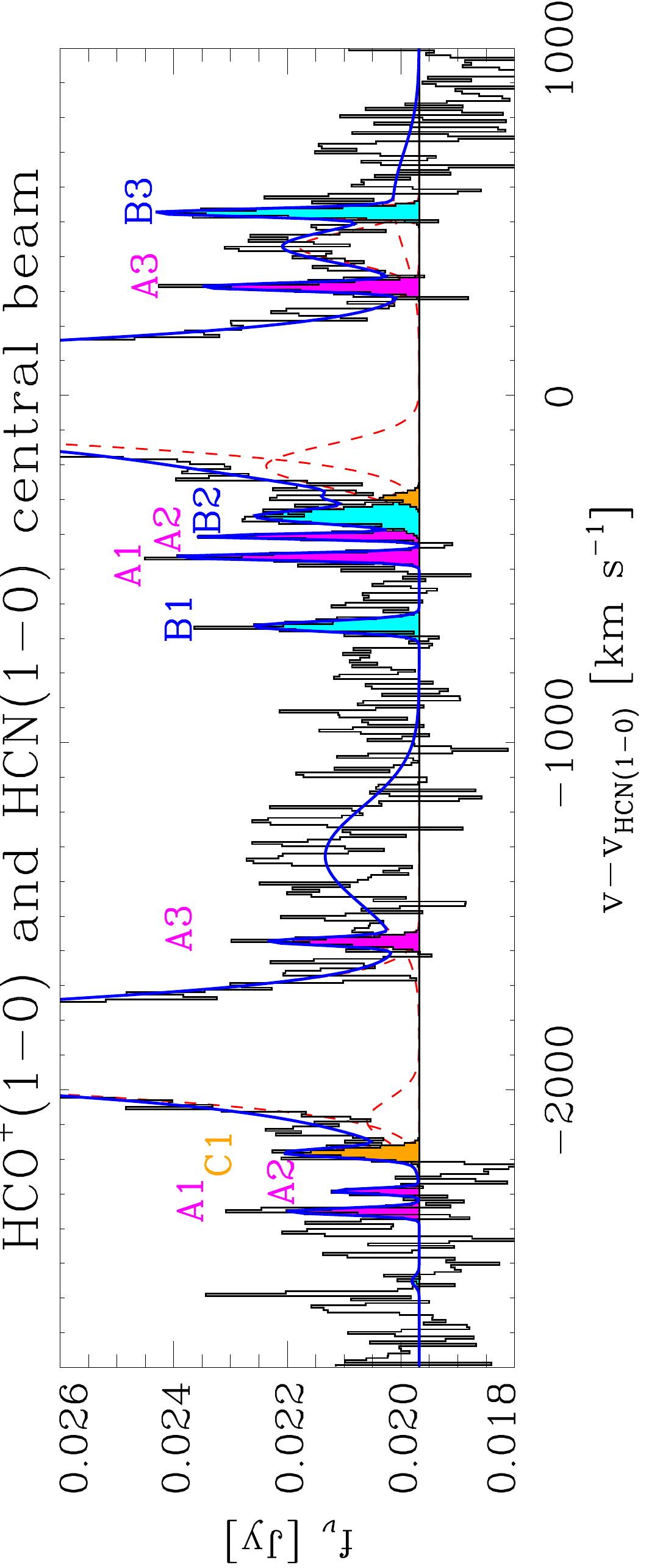}\\
	\caption{Spectral fit to the HCN(1-0) and HCO$^+$(1-0)
	line wings showing the presence of several spectrally-resolved narrow substructures (here often indicated as `clumps'). The spectrum was extracted from the cleaned datacube before continuum-subtraction and using a beam-size aperture centred at RA(J2000)=12:56:14.216 and Dec(J2000)=56:52:25.186. The rms is 1.8~mJy per $\delta v=6.9$~\kms~spectral channel. The clumps' solutions that were simultaneously fitted to both transitions are plotted in magenta, whereas features detected only in HCN (HCO$^+$) are shown in cyan (orange). The best-fit Gaussian parameters are reported in Table~\ref{table:clumps}.}\label{fig:clumps_sim}
\end{figure*}

\begin{figure*}[tb]
	\centering
	\includegraphics[clip=true,trim=0cm 0.cm 0cm 0cm,scale=.27,angle=270]{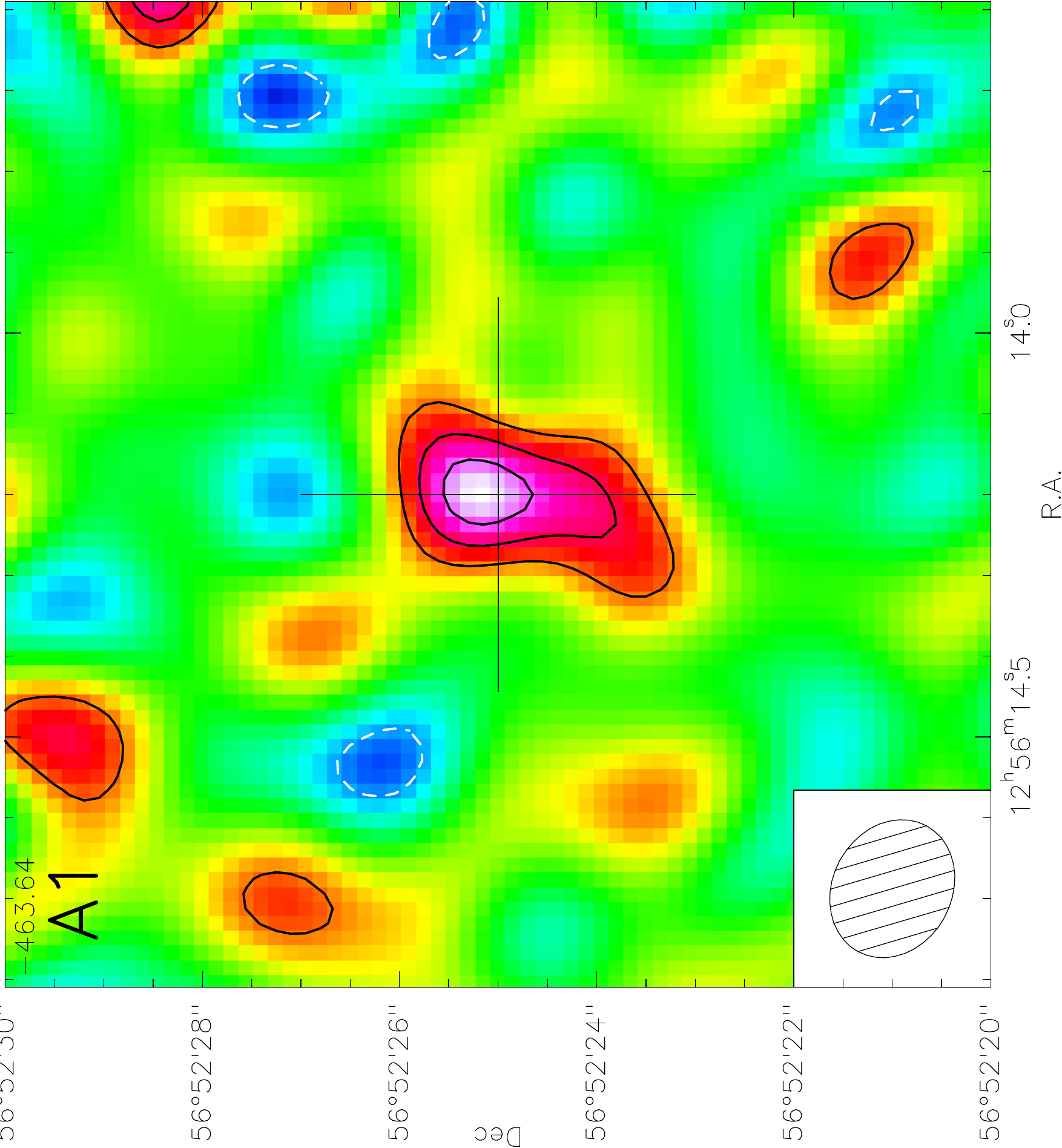}\quad
	\includegraphics[clip=true,trim=0cm 0.cm 0cm 0cm,scale=.27,angle=270]{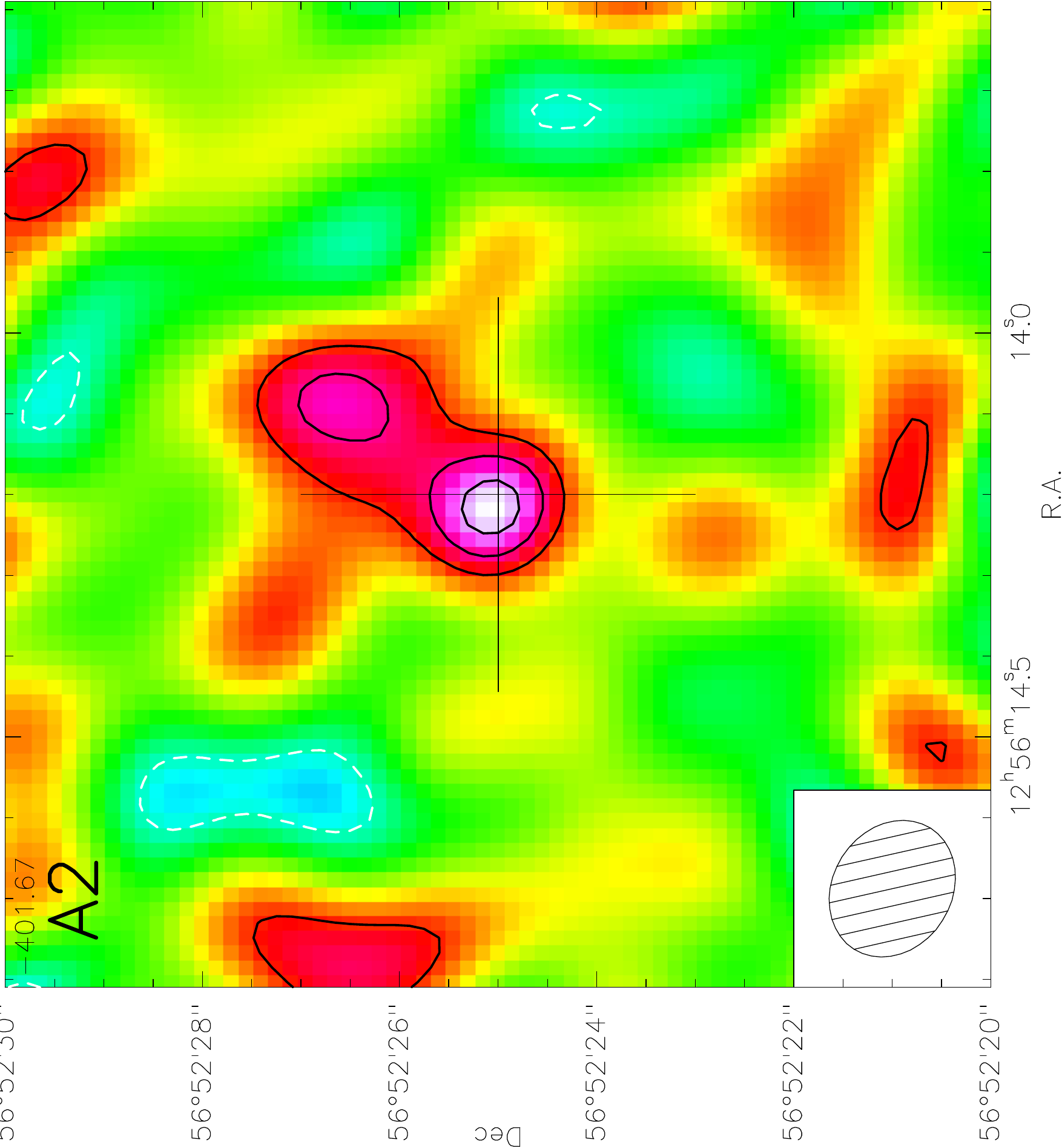}\quad
	\includegraphics[clip=true,trim=0cm 0.cm 0cm 0cm,scale=.27,angle=270]{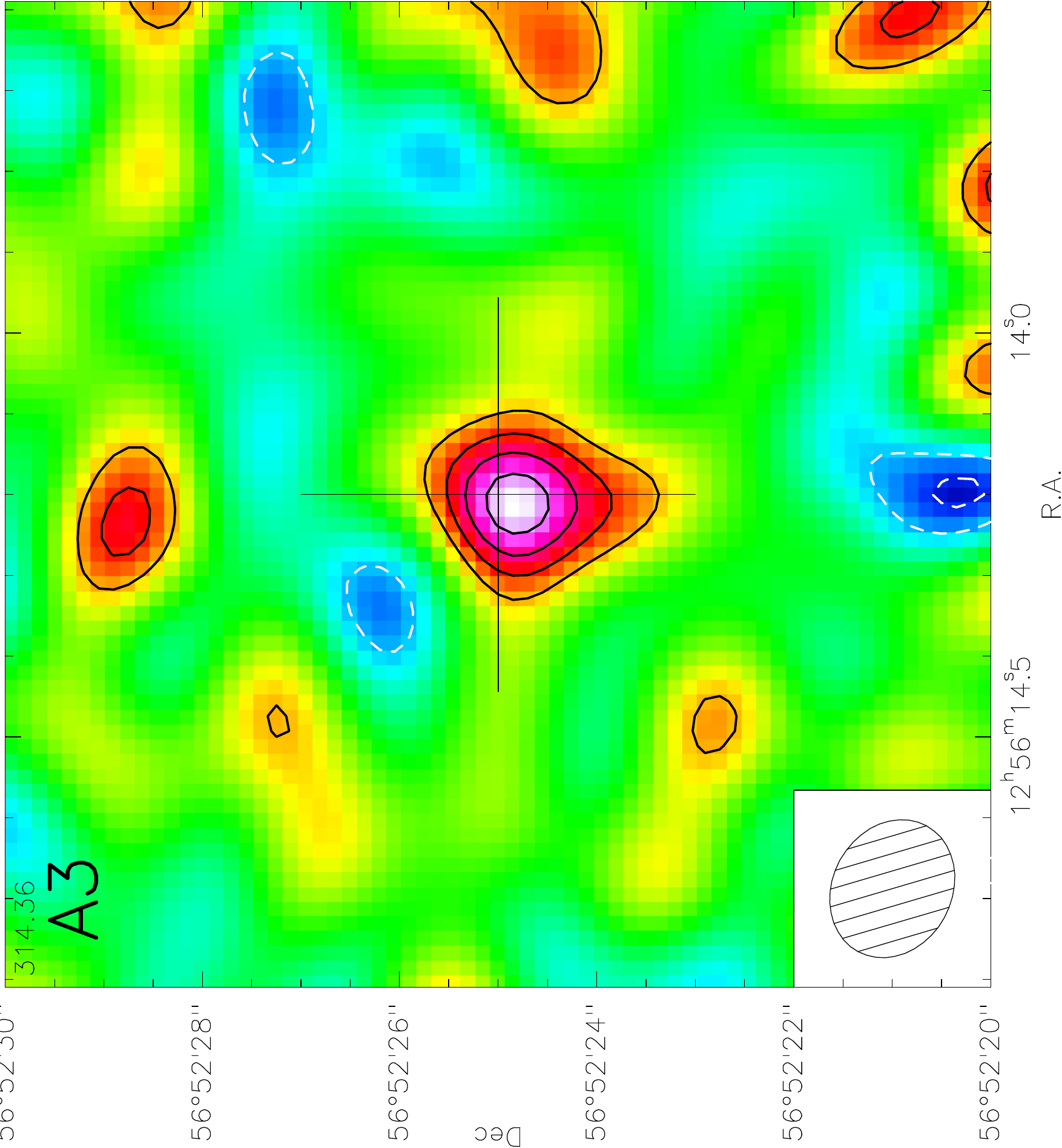}\\
	\includegraphics[clip=true,trim=0cm 0cm 0cm .3cm,scale=.27,angle=270]{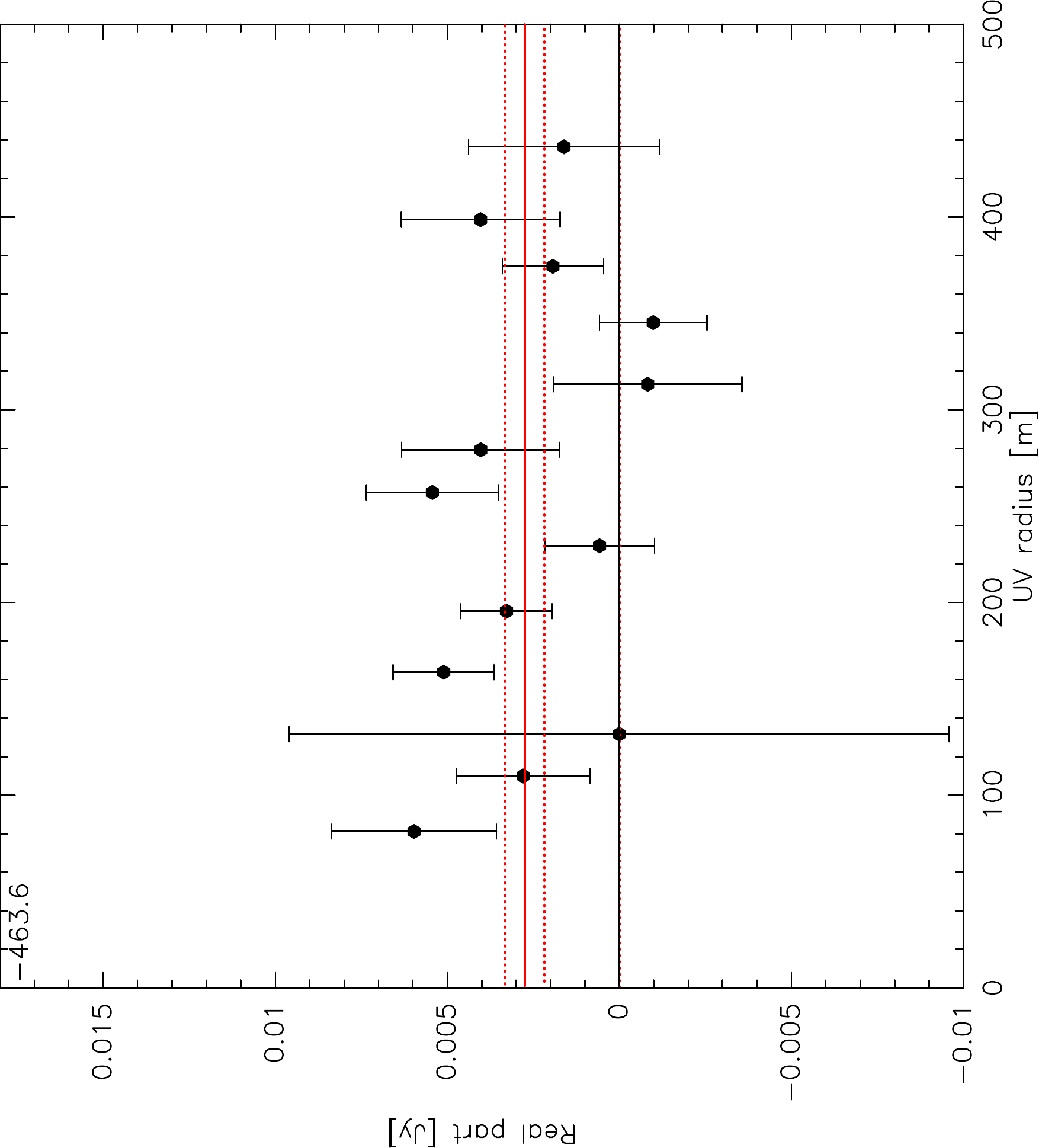}\quad
	\includegraphics[clip=true,trim=0cm 0cm 0cm .3cm,scale=.27,angle=270]{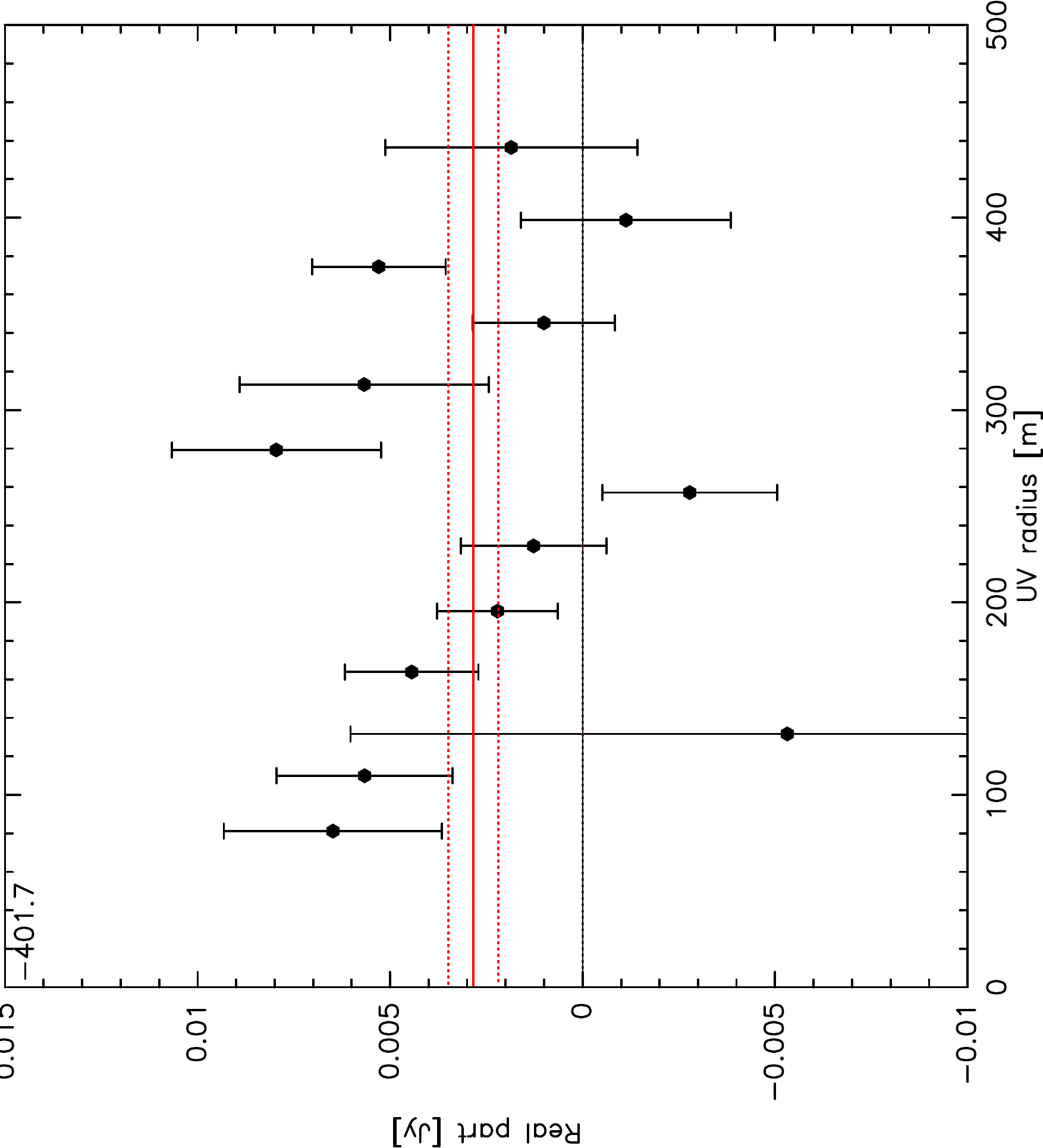}\quad
	\includegraphics[clip=true,trim=0cm 0.cm 0cm .3cm,scale=.27,angle=270]{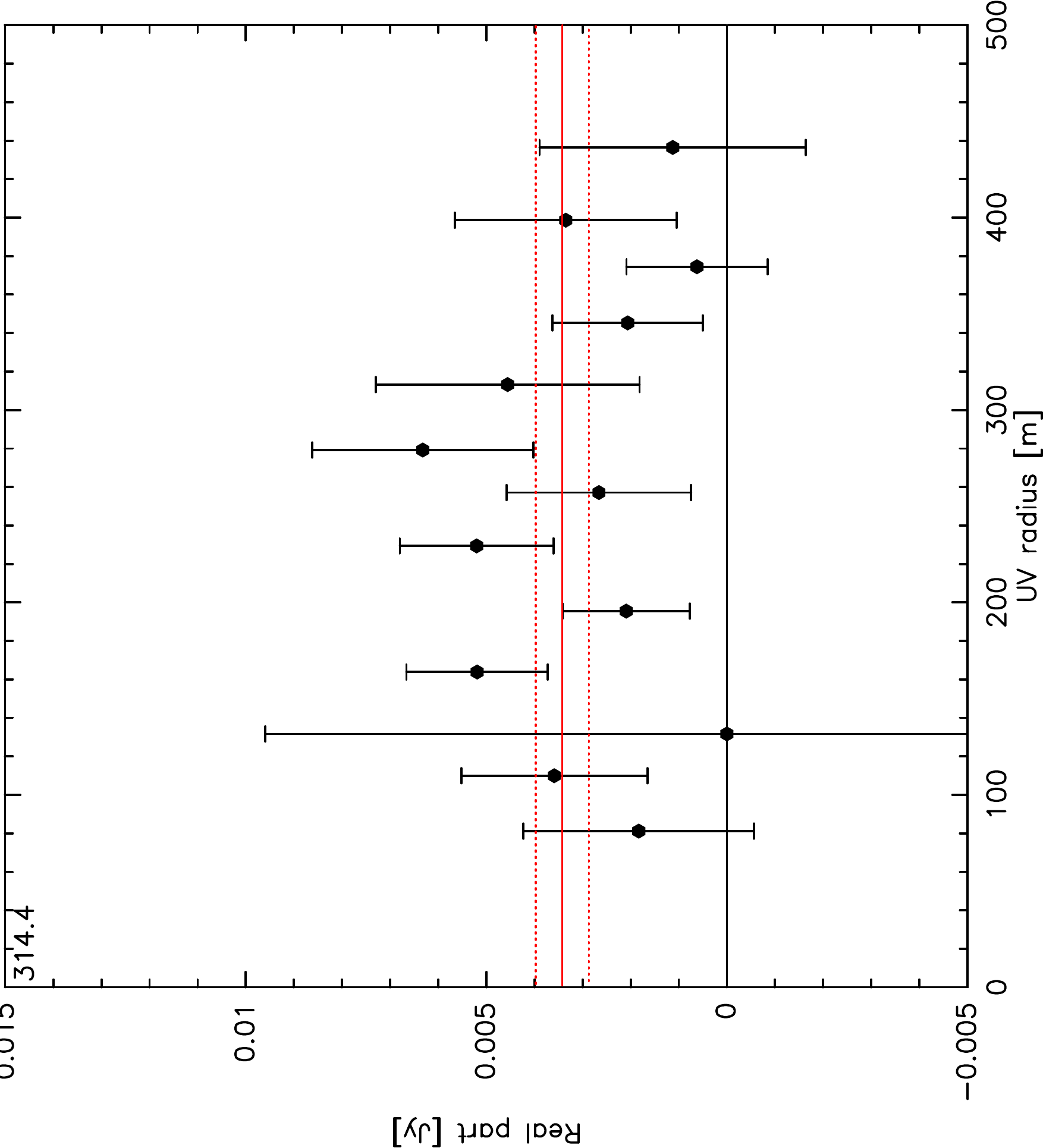}\\
	\includegraphics[clip=true,trim=0cm 0.cm 0cm 0cm,scale=.27,angle=270]{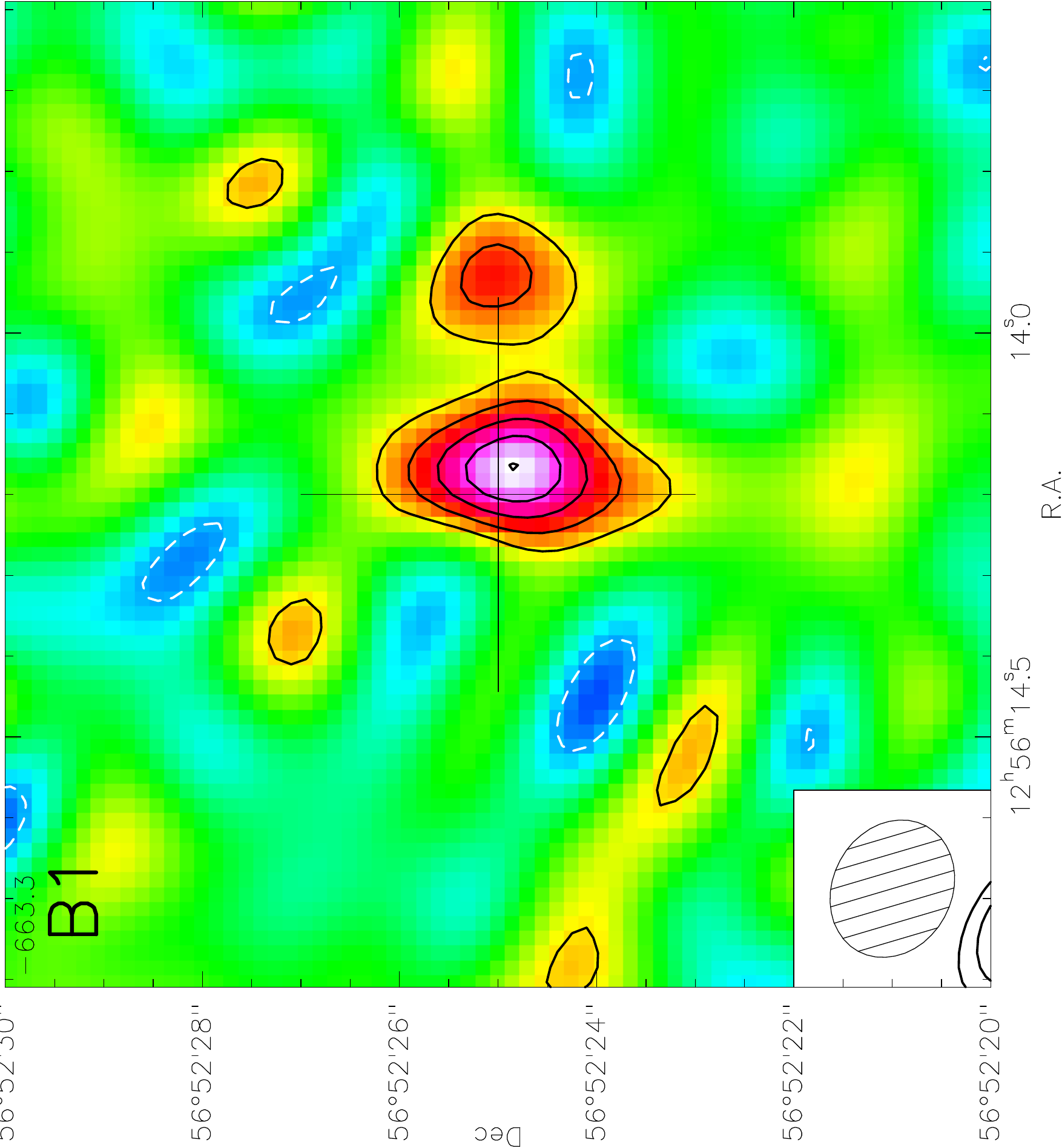}\quad
	\includegraphics[clip=true,trim=0cm 0.cm 0cm 0cm,scale=.27,angle=270]{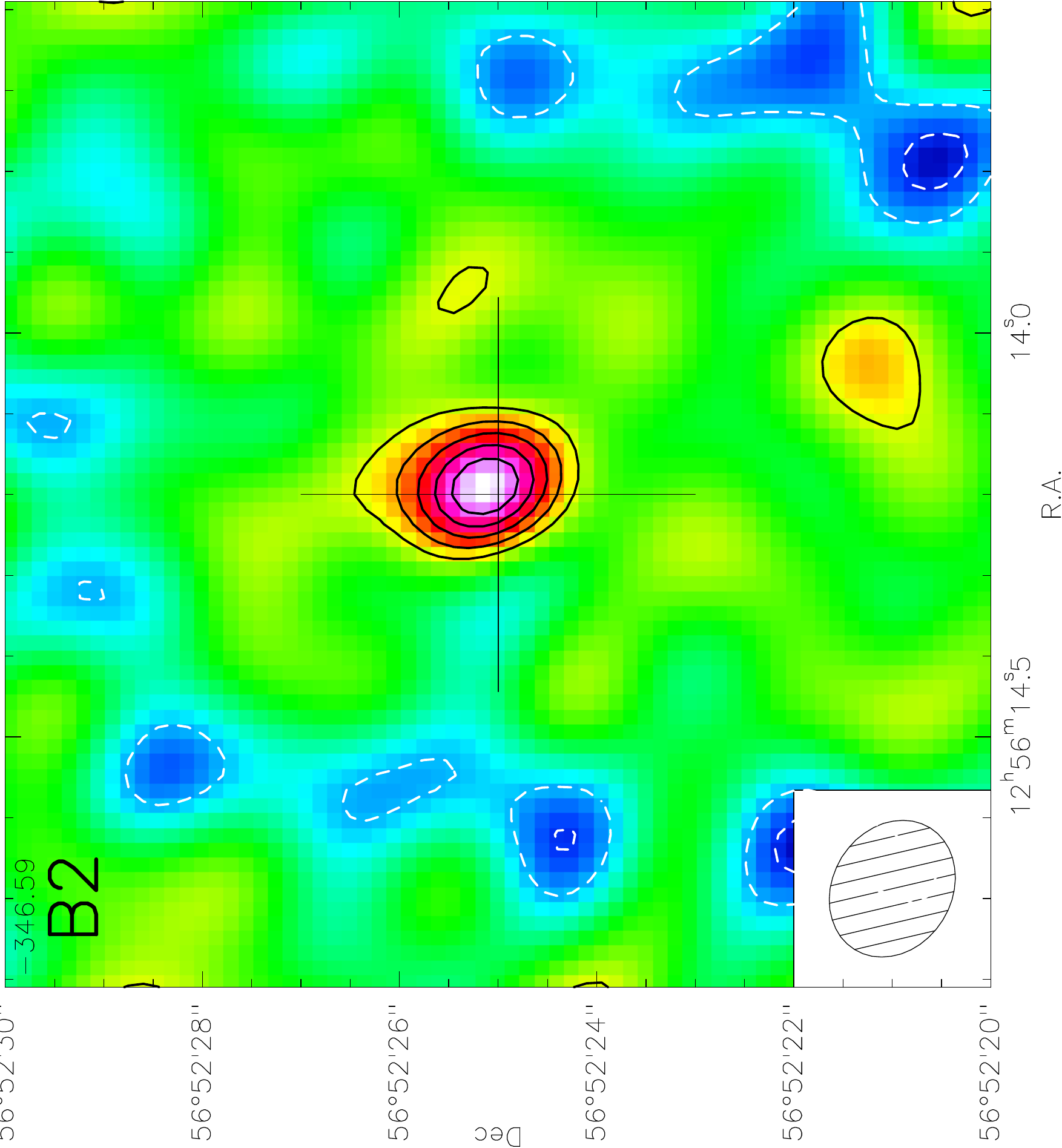}\quad
	\includegraphics[clip=true,trim=0cm 0.cm 0cm 0cm,scale=.27,angle=270]{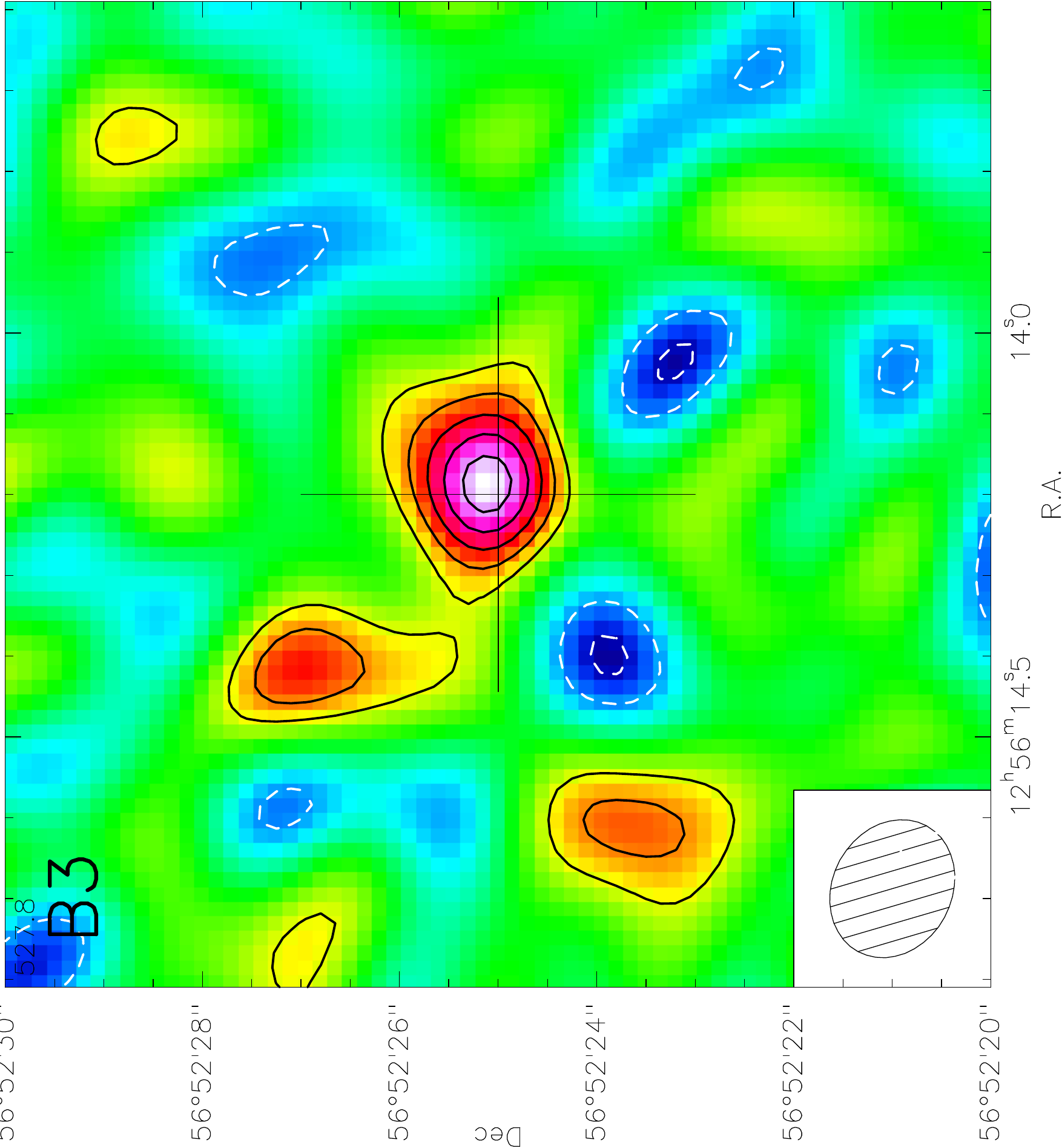}\\
	\includegraphics[clip=true,trim=0cm 0.cm 0cm .3cm,scale=.27,angle=270]{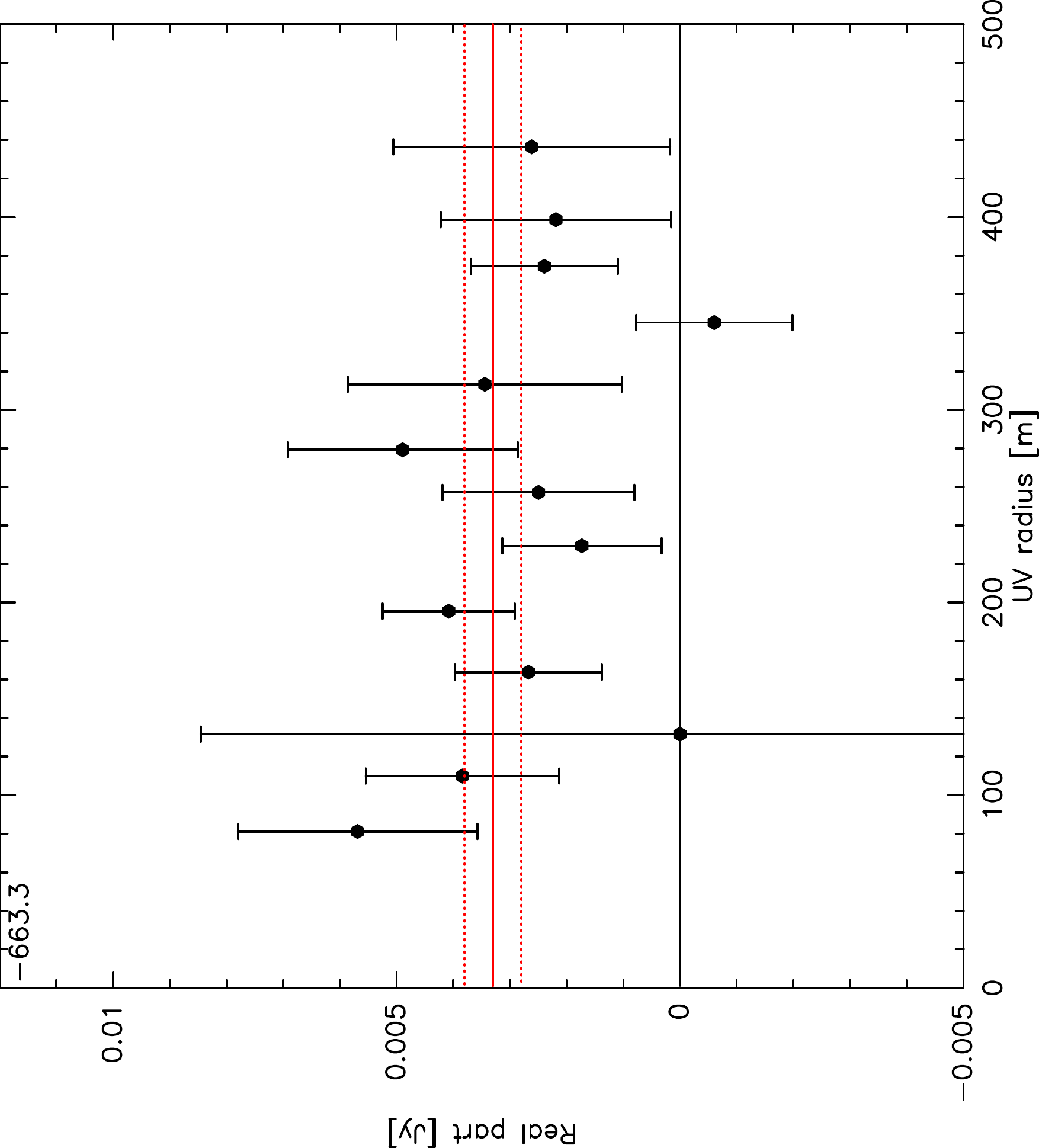}\quad
	\includegraphics[clip=true,trim=0cm 0.cm 0cm .3cm,scale=.27,angle=270]{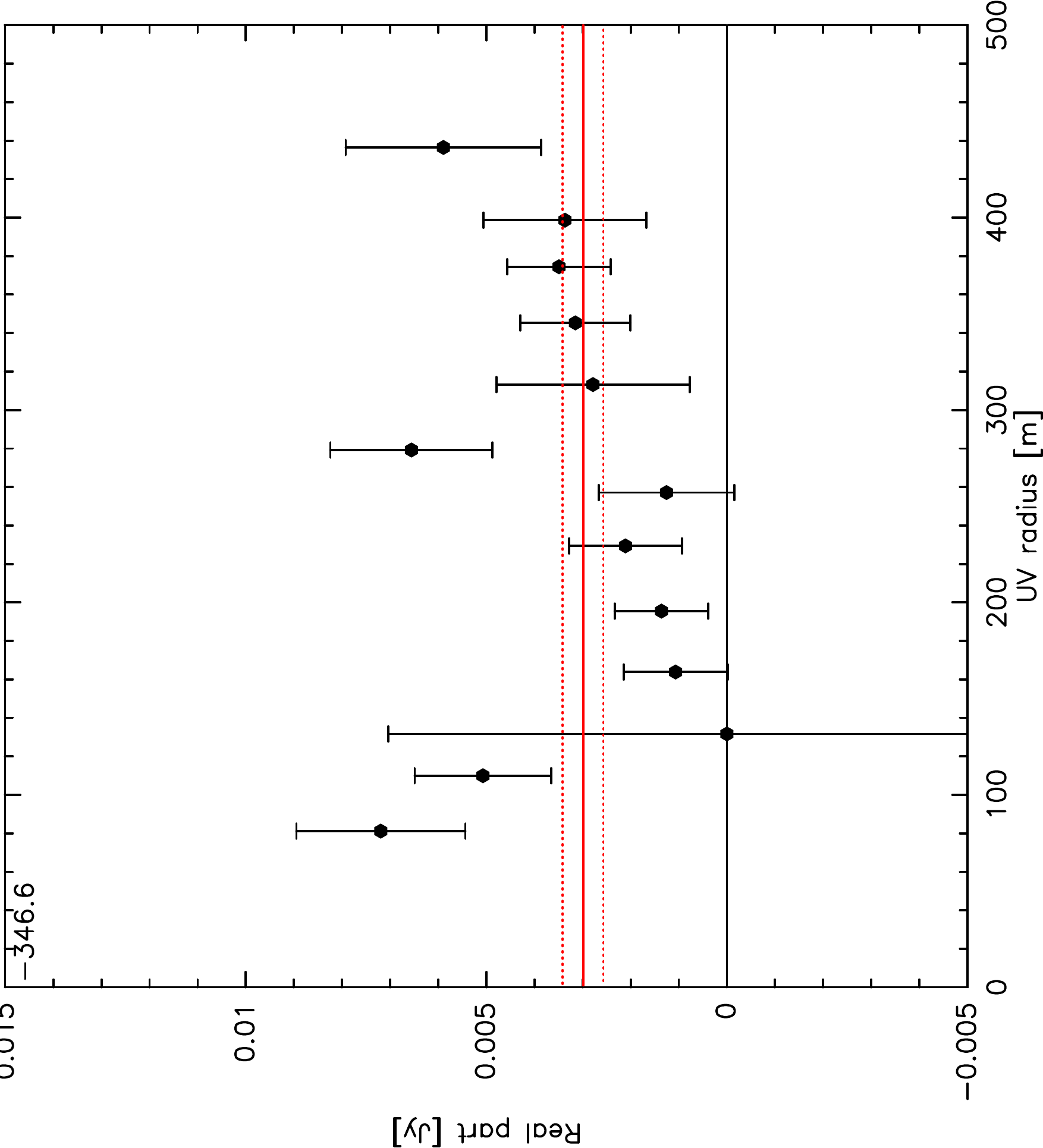}\quad
	\includegraphics[clip=true,trim=0cm 0.cm 0cm .3cm,scale=.27,angle=270]{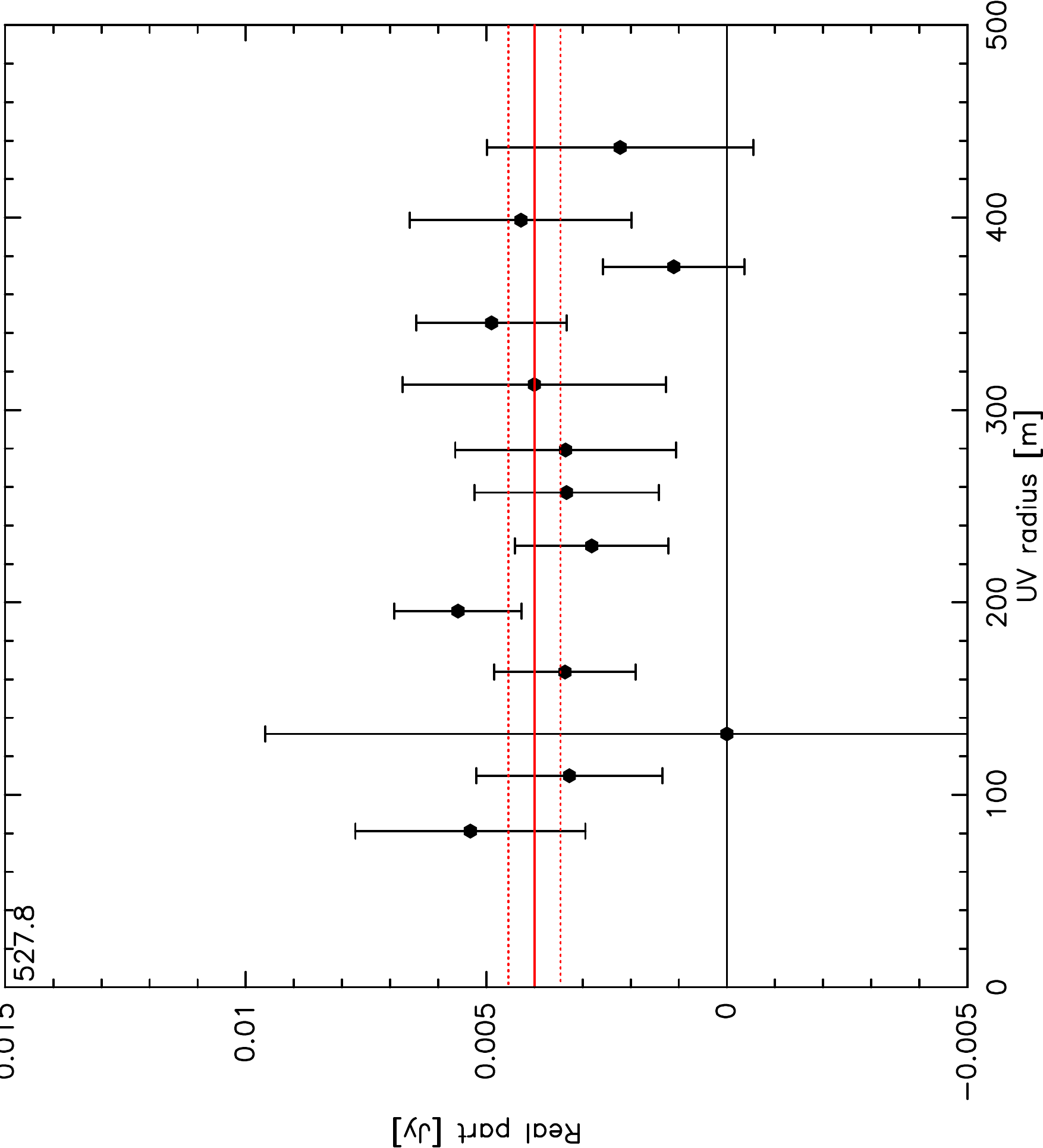}\\
	\caption{Interferometric maps and corresponding {\it uv} visibility plots of the HCN(1-0) clumps (top rows: A1, A2, A3; bottom rows: B1, B2, and B3, see also labels on the top-left corner of the maps). The maps show the continuum-subtracted emission integrated between $v_c-2\sigma_v<v<v_c+2\sigma_v$, where $v_c$ and $\sigma_v$ are the best-fit central velocity and velocity dispersion obtained from the spectral line fitting (Table~\ref{table:clumps}). Each map is $10''\times 10''$, and the synthesised beam is $1.45'' \times 1.21''$. Contours correspond to ($-3\sigma$, $-2\sigma$, $2\sigma$, $3\sigma$, $4\sigma$, $5\sigma$, $6\sigma$). Below each map we report the corresponding {\it uv} plot, binned in intervals of 30m. The red solid and dashed lines represent the best-fit value and its associated uncertainty obtained using a point source model (Table~\ref{table:clumps}).  
}\label{fig:clumps_maps}
\end{figure*}

\begin{table*}[tbp]
	\centering \small
	\caption{Analysis of the HCN(1-0) and HCO$^+$(1-0) outflowing clumps}
	\label{table:clumps}
	\begin{tabular}{lccccccccccc}
		\hline
		\hline
		ID	& \multicolumn{7}{c}{Spectral fit (Gaussian)} 			& \multicolumn{4}{c}{{\it uv} visibility fit (point source model)} \\	
		\hline
		 & $v$ & $\sigma_v$ & $S_{\rm peak}^{\rm HCO+}$ & $S_{\rm peak}^{\rm HCN}$ & S/N & $\int S^{\rm HCN}dv$ & $M_{\rm mol}^{\ddag}$ & $\rm \delta RA$ & $\rm\delta Dec$ & S/N & $\int S^{\rm HCN} dv$ \\		
		 & [\kms]  & [\kms] & [mJy]  & [mJy] & (HCN) & [Jy~\kms] & [$10^8~M_{\odot}$] & [$\arcsec$] & [$\arcsec$] &	& [Jy~\kms] \\
		\hline			   
		\hline
		A1 & $-465\pm2$ & $9\pm2$ & $2.4\pm0.8$ & $4.4\pm0.9$ & 4.9 &   $0.10\pm0.03$ & 0.5\text{-}3.6 & $-0.05\pm0.13$ & $0.17\pm0.12$   &   4.8  & $0.11\pm0.02$   \\	
		A2 & $-406\pm2$ & $7\pm2$ & $1.6\pm0.9$ & $3.9\pm1.0$ & 3.9 &   $0.07\pm0.03$ & 0.3\text{-}2.4 & $0.03\pm0.15$  &  $0.09\pm0.13$  &  4.4  & $0.08\pm0.02$ \\
		A3 & $313\pm3$	& $11\pm3$ & $2.2\pm0.8$ & $3.6\pm0.8$ & 4.5 &  $0.10\pm0.03$ & 0.4\text{-}3.4 & $0.11\pm0.10$  &  $-0.23\pm0.09$  & 6.2   &  $0.15\pm0.03$  \\
		\hline
		B1 & $-665\pm4$ & $14\pm4$ & $<2.1^{\dag}$ & $3.0\pm0.7$ & 4.3 &  $0.11\pm0.04$ & 0.5\text{-}3.7 &  $-0.28\pm0.10$  &  $-0.16\pm0.09$  &  6.6  &  $0.17\pm0.03$   \\
		B2 & $-349\pm6$ & $20\pm6$ & $<2.1^{\dag}$ & $2.8\pm0.7$ & 4.0 &  $0.14\pm0.05$ & 0.6\text{-}4.9 & $-0.12\pm0.09$  & $0.12\pm0.08$  & 7.1  &  $0.24\pm0.03$   \\
		B3 & $526\pm3$ & $11\pm3$ & $<2.7^{\dag}$ & $4.1\pm0.9$ & 4.6 &  $0.11\pm0.04$ & 0.5\text{-}3.9 &  $-0.11\pm0.09$ &  $0.10\pm0.08$  &  7.4  & $0.16\pm0.02$  \\
		\hline
		C1 & $-297\pm6$  &  $13\pm6$   & $2.1\pm0.8$    &  $<2.4^{\dag}$ & $<3$ & $<0.07^{\dag}$ & $<2.0^{\dag}$ &  &    &   &  \\
		\hline
	\end{tabular}
	
	\begin{flushleft}
		$^{\dag}$ 3$\sigma$ upper limit;
		$^{\ddag}$ Range of dense molecular gas masses estimated from the HCN(1-0) line luminosity following $M_{\rm mol}~[M_{\odot}]=\alpha_{\rm HCN}~L^{\prime}_{\rm HCN(1-0)}~[\rm K~km~s^{-1}~pc^2]$, assuming $\alpha_{\rm HCN} \in [3.1, 24]~\rm M_{\odot}~(K~km~s^{-1}~pc^2)^{-1}$, where the lower $\alpha_{\rm HCN}$ value has been proposed by \cite{Garcia-Burillo+12} for (U)LIRGs, whereas the higher $\alpha_{\rm HCN}$ has been measured by \cite{Leroy+15} for individual GMCs in the starburst region of NGC~253. We note that both \cite{Gao+Solomon04}, based on integrated observations of local galaxies and on the assumption that HCN traces very dense gas, and \cite{Shimajiri+17}, based on observations of Galactic GMCs (see also \cite{Kauffmann+17}) propose an intermediate value of $\alpha_{\rm HCN} = 10~\rm M_{\odot}~(K~km~s^{-1}~pc^2)^{-1}$.
	\end{flushleft}
\end{table*}

The broad wings of the HCN(1-0) and HCO$^+$(1-0) lines, first detected by \cite{Aalto+12}, show several narrow substructures when examined at high spectral resolution. These features are clearly visible in the spectrum extracted
from the central beam (containing $\sim70$\% of the total HCN(1-0) flux), presented in Figure~\ref{fig:clumps_sim}. 
To analyse the properties of these spectral clumps, we performed a simultaneous fit to the HCN(1-0) and HCO$^+$(1-0) transitions by using multiple Gaussian components, whose results are overlaid on the data in Fig~\ref{fig:clumps_sim}.
The fit confirms that clumps labelled as `A1', `A2', and `A3' are detected in both the HCN and HCO$^+$ lines (with the same velocity and dispersion), whereas the HCN clumps labelled as `B1', `B2', and `B3' do not have a counterpart in HCO$^+$.
Only one feature, labelled as `C1', is detected in HCO$^+$ but not in HCN. As discussed later on, all these clumps are detected at high significance.
Such differences between those species were already pointed out by \cite{Lindberg+16} based on the analysis of Mrk~231's spectral line profiles. In particular, they noted the presence of HCN and HCO$^+$ features at different velocities in the line wings. Moreover, their radiative transfer modelling of the HCN and HCO$^+$ line emission is inconsistent when using the same outflow properties for both species: while a clumpy outflow model could reproduce the HCN emission, the same model would lead to an unrealistically high outflow mass for HCO$^+$. These facts suggest that the HCN and HCO$^+$ line emissions arise in regions with somehow different physical/chemical properties, hereby a clumpy and chemically differentiated outflow. One possible origin of this difference could be related to the presence of shock fronts in the outflow, with HCN tracing the shocked regions between the outflow and the surrounding gas, and high-velocity HCO$^+$ tracing pre-shock gas, closer to the nucleus.

The best-fit Gaussian parameters of the clumps are reported in Table~\ref{table:clumps}. The velocity dispersions of $\sigma_v\sim7-20$~\kms are consistent with individual GMCs. These $\sigma_v$ values are larger than typical Milky Way GMCs ($\langle \sigma_v\rangle \sim2$~\kms) but smaller than those measured for clouds populating the dense nuclear starburst of NGC~253 ($\langle \sigma_v\rangle \sim22$~\kms, \citealt{Leroy+15}). By assuming HCN-to-H$_2$ conversion factors in the range $3.1-24~\rm M_{\odot}~(K~km~s^{-1}~pc^2)^{-1}$ (e.g. \citealt{Gao+Solomon04, Aalto+15, Leroy+15}), the HCN luminosities of the clumps would correspond to molecular gas masses of $M_{mol}\sim 0.3-4.9\times10^8~M_{\odot}$ (see Table~\ref{table:clumps}). Such masses are one order of magnitude larger than typical GMCs in the MW and M33 \citep{Rosolowsky+03}, but roughly comparable (in the lower-mass end limit) to the more massive GMCs populating the dense environment within the nuclear starbust of NGC~253 \citep{Leroy+15}.  

The spectral clumps appear less significant in spectra extracted from apertures larger than the central beam. Only features A1, A2, B2 and B3 are clearly detected also in the spectrum extracted from a $4''$-size aperture, with best-fit $v$ and $\sigma_v$ values consistent with those measured within the central beam (Table~\ref{table:clumps}). The lower significance of the spectral clumps in large apertures is due to the fact that they are spatially unresolved at the resolution of the HCN(1-0) data ($\sim1$~kpc, see Table~\ref{table:obs}). Hence, their flux is maximised in the central beam, but it is washed out in larger apertures because of the superposition with additional outflow components. 

We checked the significance of the HCN(1-0) clumps in the $uv$ visibility data, by analysing the $uv$ amplitude as a function of $uv$ radius ({\it uv} plots). The interferometric images of the HCN(1-0) clumps and the corresponding {\it uv} plots are shown in Fig.~\ref{fig:clumps_maps}. Both the maps and the $uv$ plots, with their flat trend of $uv$ visibilities as a function of $uv$ radius, confirm that the HCN clumps are spatially unresolved in our observations. A fit to the $uv$ plots using a point source model, whose results are also listed in Table~\ref{table:clumps}, shows that the HCN clumps (A1-A3, B1-B3) are individually detected at a high significance, ranging from $S/N\sim 4.8-7.4$. Furthermore, as shown in Table~\ref{table:clumps}, the integrated fluxes derived from the $uv$ fit are overall consistent with those obtained from the simultaneous spectral fit shown in Figure~\ref{fig:clumps_sim}. Since the $uv$ visibility data contain the {\it total} flux within a certain velocity range (chosen as $v_c\pm2\sigma_v$, where $v_c$ and $\sigma_c$ are the central velocity and dispersion of each spectral clump, listed in Table~\ref{table:clumps}), while the spectrum in Fig.~\ref{fig:clumps_sim} samples only the central beam, such flux consistency between the central beam and the $uv$ data suggests that the clumps dominate the HCN flux within the velocity ranges sampled by their narrow line profile. In other words, at the velocity of the clumps, there is little additional flux in more spatially extended and diffuse components. 

We searched for the presence of clumps also in the other molecular gas tracers, by inspecting the spectra extracted from the central beam. The HNC(1-0) and CN(1-0) line wings show clumpy features similar to HCN(1-0), but the lower S/N of these datasets and, in the case of CN, the blending of the two spin-groups, do not allow us to reliably fit any of these features. In the CO(1-0) transition, the line wings are intrinsically much smoother compared to the high density tracers, even when studied at the highest available spectral resolution. The only discernible structure in the CO(1-0) line wings is a red-shifted feature with best-fit Gaussian parameters: $v=459\pm10$~\kms, $\sigma_v = 39\pm17$~\kms and $S_{\rm CO}^{peak}=1.7\pm0.6$~mJy, which does not correspond to any of the clumps revealed in HCN or HCO$^+$.

Our results support a scenario where the molecular outflow entrains high-velocity, dense molecular gas clouds, dominating the emission in the HCN and HCO$^+$ line wings, which are embedded in a more diffuse phase that contributes most to the observed CO(1-0) line wings. A similar conclusion was independently reached by \cite{Cicone+18} for the dual AGN and LIRG NGC~6240, based on a multi-tracer analysis of the extended molecular outflow.

\section{Discussion}\label{sec:discussion}

\subsection{Enhanced CN/HCN abundance ratios} 

We aim at understanding the origin of the enhanced CN/CO(1-0) and HCN/CO(1-0) line luminosity ratios measured in the outflow of Mrk~231 (Figure~\ref{fig:line_ratio_plots}). This can be due to a combination of (i) higher fractional abundances of such tracers ($X_{\rm HCN}$, $X_{\rm CN}$, i.e. the abundance ratios of these species over the total H$_2$ abundance) {\it and/or} (ii) a substantial entrainment of dense gas with $n>10^4$~cm$^{-3}$, which would increase the excitation of these lines \citep{Aalto+12a,Aalto+15}. A high abundance of HCN may be the signature of shocks (e.g. see \citealt{Aalto+12a} and references therein), whereas a high $X_{\rm CN}$ should trace gas exposed to a strong UV radiation field (PDR-like behaviour) from massive stars. \cite{Garcia-Burillo+10} proposed that CN emission is amplified in the vicinity of an AGN due to the X-ray radiation field, although there is still no consensus on this result, since many AGN hosts show the opposite behaviour, with $L^{\prime}_{\rm CN}/L^{\prime}_{\rm CO}$ values that are lower by a factor of 2-3 compared to the average ratios measured in pure starbursts \citep{Wilson18, JP+07}. The prominent CN line wings of Mrk~231 may indicate that the outflow is bathed in a strong UV radiation field, either due to star formation in the galaxy disk or to new stars forming within the outflow itself. The latter is a scenario proposed by various recent models and simulations \citep{Zubovas+13, Zubovas+King14, Ishibashi+Fabian12, Ishibashi+13, El-Badry+16, Wang+Loeb18, Decataldo+19}. Star formation in the outflow could have major implications, since such stars would have highly radial orbit and could, for instance, contribute to the formation of the spheroidal component of galaxies. Observational confirmations of this new mode of star formation have been presented recently \citep{Maiolino+17, Gallagher+19, RodriguezDelPino+19}. Within the context of our findings, young massive stars formed inside the outflow would provide a much stronger (internal) UV radiation field with respect to the illumination form stars in the disk or even from the AGN.

By using the \texttt{RADEX}\footnote{\texttt{RADEX} is a publicly available non-local thermodynamic equilibrium radiative transfer code, which  uses as input parameters the molecular gas column density ($N_{\rm H_2}$), the line width ($\Delta v$), the gas temperature ($T_{\rm kin}$), the background temperature (set equal to the Cosmic Microwave Background temperature at $z=0$, i.e. $T_{\rm bg}=2.73$~K), and the gas volume density \citep{vanderTak+07}} dense cloud models developed by \cite{Aalto+15} to reproduce the HCN(3-2)/(1-0) line luminosity ratios in Mrk~231's outflow, we can attempt to find a combination of $X_{\rm HCN}$, $X_{\rm CN}$, $T_{\rm kin}$ and $n_{\rm H_2}$ solutions that can fit also the CN/HCN and CN spin doublet line ratios (Table~\ref{table:fit_total}). We assume that the HCN and CN line emissions arise from the same dense cloud population, while the low-J CO line emission is due to a different, more diffuse phase of the outflow. We recall that in these models (see also \citealt{Aalto+15}), the dense clouds can be either self gravitating, virialized clouds, i.e. implying that their internal velocity dispersion ($\Delta v_{\rm sg}$) is locked to their mass ($M_{\rm vir}$) and size ($R$) through $\Delta v_{\rm sg}=(GM_{\rm vir}/G)^{1/2}$, or unbound clouds, for which $\Delta v\gg\Delta v_{\rm sg}$. We explore CN and HCN abundances in the range between $10^{-8}$ and $10^{-6}$.
We find that, depending on whether the clouds are self gravitating or unbound, the models produce very different values for the absolute CN and HCN abundances, hence $\rm X_{CN}$ and $\rm X_{HCN}$ remain quantitatively unconstrained for the outflow with current data. However, all possible solutions that fit the observed line ratios consistently require $\rm X_{CN}>X_{HCN}$, with an CN abundance that is at least a factor of three higher than the HCN abundance. Gas densities for such outflow phase (traced by the CN and HCN emissions) are $\rm n_{H_2}\sim10^5-10^6$~cm$^{-3}$, with temperatures not much higher than $\rm T_{kin}\sim20$~K. Since CN is a well-known PDR tracer (see also $\S$~\ref{sec:introduction}), these results strongly suggest that the whole dense cloud population in outflow is affected by UV radiation. We should mention that high CN abundances may be due also to cosmic rays (e.g. see work done on the Galactic centre by \cite{Harada+15}), which are known to permeate Mrk~231's outflow as inferred by \cite{Gonzalez-Alfonso+18} based on the OH$^+$ enhancement. However, it is not clear whether a cosmic ray chemistry would also explain $\rm X_{CN}>X_{HCN}$.

In the literature (see Appendix~\ref{sec:AppendixB}), there are only two other extragalactic molecular outflows with available CN(1-0) observations (NGC~3256 and NGC~253, see also Table~\ref{table:literature}), and neither of them shows the exceptionally high $L^{\prime}_{\rm CN(1-0)}/L^{\prime}_{\rm CO(1-0)}$ values that we measure in Mrk~231. However, in NGC~3256, \cite{Sakamoto+14} and \cite{Harada+18} find an enhancement of CN and HCN at the outflow position compared to other locations within the galaxy merger. 
Different from CN, there are quite a few extragalactic outflows with HCN and/or HCO$^+$ observations. In Arp~220, observations by \cite{BarcosMunoz+18} imply
$L^{\prime}_{\rm HCN(1-0)}/L^{\prime}_{\rm CO(1-0)}\sim0.9$ and $\sim7$ for the blue- and red-shifted outflows, which are significantly higher than the corresponding luminosity ratio of $0.20\pm0.03$ obtained by using the HCN and CO emissions integrated across the whole galaxy (Table~\ref{table:literature}). In IC~5063, the high HCO$^+$(4-3)/CO(2-1) ratio measured in the outflow by \cite{Oosterloo+17} points to an additional dense component with $\rm n\gtrsim 10^6~cm^{-3}$ that is not accounted for by the low-J CO emission (which is likely optically thin, e.g. \cite{Dasyra+16}). 
\cite{Aladro+18} detected the blue-shifted component of Mrk~273's molecular outflow in HCN(3-2) emission, and its location is consistent with the CO(1-0) component imaged by \cite{Cicone+14}. Among other extragalactic outflows detected in dense molecular gas tracers we list NGC~1068 (HCN(4-3) and HCO$^+$(4-3) emission, \citealt{Garcia-Burillo+14}), NGC~1266 (CS(2-1) and HCN(1-0) emission, \citealt{Alatalo+15}), and IC~860 (CS(7-6) absorption, \citealt{Aalto+19}).

Interestingly, even though the molecular streamers embedded in the starburst-driven outflows of NGC~253 and M~82 have been detected in dense gas tracers, the latter do not show a significant enhancement compared to CO emission. In NGC~253, the CN emission is boosted (with respect to both CO and C$^{17}$O) in the central starburst, but not in the streamers \citep{Meier+15}. The HCN/CO luminosity ratio is instead similar between starburst and outflow \citep{Walter+17}. 
The outflow of M82 was detected in HCO$^+$(1-0) and HCN(1-0) emission \citep{Salas+14}, but the mass traced by these molecules is a small fraction ($\gtrsim2$\%) of the total (CO-based) mass of the outflow. Hence, the molecular gas filaments embedded in the starburst-driven outflows of NGC~253 and M82 appear to differ in their physical properties from Mrk~231's outflow. 

The exceptional boost of CN and HCN in Mrk~231's outflow may hold clues to explain the enhancement of such dense gas tracers in integrated observations of this ULIRG. Indeed, although the total flux is dominated by the narrow component of the molecular lines (as also reflected by the line ratios in Fig.~\ref{fig:line_ratio_plots}, where the integrated data point lies closer to the core than to the wings) we note that the CN/CO and HCN/CO luminosity ratios measured in the narrow core are quite high compared to the nearby galaxy population and to local GMCs, although not as extreme as those measured in the wings. We cannot rule out that the low-projected velocity component of the outflow is contaminating significantly also the narrow emission\footnote{Only very high S/N, spatially resolved observations would allow us to quantify such contribution, by de-blending the low-velocity outflow emission from the rotating disk.}. In this case, since such component would share the same CN and HCN enhancement of the high velocity wings, it would also boost the line ratios measured for the narrow core. In the hypothesis that Mrk~231-like outflows have a chemical impact on the global ISM of their host galaxies, they would resemble scaled-up versions of the so-called `chemically active' protostellar outflows revealed in young stellar objects (YSOs), which are characterised by enhanced abundances of shock sensitive tracers such as SiO \citep{Tafalla+Hacar13} or even CN and HCO$^+$ \citep{Bachiller+01}.

\subsection{Dense clumps embedded in the outflow}

The discovery of narrow features in the HCN and HCO$^+$ line wings suggests that the outflow embeds coherent dense molecular gas complexes, with velocity dispersions similar to extragalactic GMCs detected in dense environments. Furthermore, our data show that, while such clumps dominate the high-velocity emission from the high density tracers, they are not present in the wings of low-J CO lines. The HCN clouds embedded in Mrk~231's outflow are massive, with inferred $M_{\rm mol}$ values in the range $\sim0.3-4.9\times10^8~M_{\odot}$ (Table~\ref{table:clumps}), and have large velocity dispersions of $\sim7-20$~\kms. Such unusually high masses may result from an observational bias, where multiple outflowing clouds with overlapping line-of-sight velocities contribute to the narrow sub-structures that we identify in the (beam-averaged) HCN spectra in Fig.~\ref{fig:clumps_sim}. However, it is also possible that we are witnessing the survival of only the biggest and/or the most massive molecular clouds in outflow, which are less affected by photoevaporation \citep{Decataldo+19} and hydrodynamical instabilities \citep{Gronke+Oh18, Armillotta+17}. 

These physical properties, i.e. masses and velocity dispersions a factor of ten larger than GMCs in the Galactic disk, seem to be more typical of molecular clouds forming in dense environments, such as those residing in the nuclear starburst of NGC~253. The latter, identified by \cite{Leroy+15} through HCN-to-CO intensity peaks (one of which may be even entrained in the outflow), are characterised by higher surface and volume densities, shorter free fall time-scales and so higher star formation efficiencies compared to typical disk GMCs. 
That the HCN clumps detected in Mrk~231's outflow correspond to star forming complexes is a very interesting prospect, especially in the context of the new scenarios predicting the formation of stars inside outflows. Indeed, dense and clumpy gas are conditions that typically result into rapid gravitational collapse and star formation \citep{Decataldo+19}; hence our finding provides further support for this new mode of star formation.

We note that, if the clouds' mass distribution in outflow differs strongly from that of typical GMCs, as suggested by our results, it would result in a different initial mass function (IMF) of the stars that would form within such environment \citep{Hopkins12}, and possibly explain other exotic properties of (U)LIRGs such as their high $^{18}$O abundance (which favours a top-heavy IMF, e.g. \citealt{Brown+19}). 
However, a valid objection is that, even if the high-velocity clumps trace individual clouds, we know that, at least in the Milky Way, not all GMCs are active stellar nurseries. Indeed, even within GMCs, HCN emission can be quite widespread and up to $\sim50$~\% of it can be associated with extended regions with $n_{\rm H_2}\lesssim10^3$~cm$^{-3}$, i.e. two orders of magnitude below the critical density for collisions with H$_2$ molecules \citep{Pety+17,Kauffmann+17,Harada+19}. In these regions, HCN may be excited by electron collisions \citep{Goldsmith+Kauffmann17}. Additional data are required to settle this issue.

\section{Summary and Conclusions}\label{sec:conclusions}
We have reported the detection of an exceptional enhancement of CN(1-0) line emission in the molecular outflow of Mrk~231. More specifically, we measured $L^{\prime}_{\rm CN(1-0)}/L^{\prime}_{\rm CO(1-0)}$ values of $0.70\pm0.13$ (blue-shifted wing) and $0.9\pm0.2$ (red-shifted wing), which are not only much higher than the value measured in the narrow core component ($0.21\pm0.03$), but also higher than any other known Galactic or extragalactic source. The CN/CO(1-0) line luminosity ratios computed for the line wings are also significantly higher than the respective HCN/CO(1-0), HCO$^+$/CO(1-0) and HNC/CO(1-0) line luminosity ratios, indicating that, among the dense gas tracers explored so far in Mrk~231's outflow, CN has the highest luminosity advantage with respect to CO. 

We have found that, in order to reproduce these ratios, the dense gas cloud population entrained in the outflow must be characterised by an CN abundance that is at least a factor of three higher than the HCN abundance, by gas densities of $n_{\rm H_2}\sim10^{5-6}$~cm$^{-3}$, and kinetic temperatures not much higher than $T_{\rm kin}\sim20$~K. This result can be interpreted (although not uniquely) in terms of an outflow chemistry that is heavily affected by UV radiation, since CN emission has been found to be amplified mainly in PDRs. A population of young massive stars formed inside the outflow, as predicted by recent models, would naturally provide an explanation for such strong UV radiation field. 

In the literature there are only two other extragalactic molecular outflows with available CN measurements, NGC~253 and NGC~3256, and neither of them shows an CN enhancement similar to Mrk~231. 
Although we are still limited by statistics, we may conclude that the outflow of Mrk~231 is rather exceptional in its chemical and physical properties. In the future, using deeper and higher spatial resolution observations, it would be interesting to explore whether the outflow component is responsible for the exceptionally strong emission from dense H$_2$ gas tracers observed in the global ISM of this ULIRG.

Additionally, we have shown that the HCN(1-0) and HCO$^{+}$(1-0) line wings are spectrally resolved into several narrow sub-structures with velocity dispersions of $\sigma_v\sim7-20$~\kms and inferred masses of $M_{mol}\sim0.3-4.9\times10^8~M_{\odot}$, which we interpret as individual dense clumps entrained in Mrk~231's outflow. These properties are consistent with the most massive extragalactic GMCs populating the nuclear starburst of NGC~253. Three of the outflowing clumps are detected simultaneously in both the HCN and HCO$^{+}$(1-0) transitions, while additional three (one) are detected only in HCN (or HCO$^+$), supporting the hypothesis of a chemically differentiated outflow. An analysis of the $uv$ visibility data indicates that these narrow HCN outflow features are spatially unresolved at the resolution of our data ($\sim1$~kpc), but, at their line-of-sight velocity, there is little additional HCN flux in more spatially extended and diffuse components. The CO spectral line wings are instead intrinsically much smoother, implying that low-J CO emission is dominated by a more diffuse component of the outflow. 

This discovery strongly suggests that the outflow of Mrk~231 embeds coherent molecular gas cloud complexes, similar to GMCs, but overall more massive (and with different chemical properties). This possibly indicates that we are witnessing the survival of only the biggest molecular clouds that are less affected by photo-evaporation and hydrodynamical instabilities, i.e. the processes that we expect to be at work in energetic multiphase outflows and that may quickly destroy most of the coldest and densest gas. If these clumps correspond to star forming complexes, as predicted by recent theoretical models, the stars that would form from gravitational collapse of these massive clouds would probably follow a different IMF. However, this intriguing hypothesis requires follow-up observations to confirm the star forming nature of such clumps and better constrain their masses.

Finally, our interferometric maps have revealed a new molecular gas structure in Mrk~231, offset to the North by $5\arcsec$. Its high CN/CO line luminosity ratio and broad blue-shifted line profiles suggest that this northern component may trace an additional extension of the outflow, reaching out to $r>5$~kpc from the nucleus.

\begin{acknowledgements}
The interpretation of our data, especially for what concerns the ISM chemistry in the presence of strong UV fields, benefited greatly from extensive discussions with the late Malcolm Walmsley. We are deeply grateful to him, not only for his help with this paper, but also for the several insightful and enlightening discussions that some of us had with him on a broad range of topics in Astrophysics.
CC thanks Jens Kauffmann for providing the Orion data, Nanase Harada and Kazushi Sakamoto for providing non-published measurements of NGC~3256, Paola Severgnini for insightful discussions about the multi-wavelength emission of Mrk~231, and Paola Caselli for valuable explanations on the properties of CN.
This work is based on observations carried out under project numbers \texttt{UA26}, \texttt{T02F}, \texttt{UB26}, \texttt{V087}, \texttt{U--D}, \texttt{W028}, \texttt{V026}, \texttt{WA85} with the IRAM NOEMA Interferometer. IRAM is supported by INSU/CNRS (France), MPG (Germany) and IGN (Spain).
This project has received funding from the European Union's Horizon 2020 research and innovation programme under the Marie Sk\l{}odowska-Curie grant agreement No 664931.
RM acknowledges ERC Advanced Grant 695671 `QUENCH’ and support by the Science and Technology Facilities Council (STFC). 
\end{acknowledgements}

\bibliographystyle{aa}
\bibliography{cn_mrk231}

\begin{appendix}

\section{A possible northern extension of the molecular outflow}\label{sec:appendixA}

The interferometric map of the total CN(1-0) line emission
shown in the left panel of Figure~\ref{fig:cn_maps} presents an extended feature, offset to the North by $\sim5\arcsec$ with respect to Mrk~231's molecular disk. As noted in the $\S$~\ref{sec:introduction}, Mrk~231 hosts multiple extranuclear shells and bubbles that are associated with feedback processes (e.g. \citealt{Lipari+05,Lipari+09,Morganti+16}). However, most of the structures identified so far are located {\it South} of the nucleus, whereas such northern CN emission does not overlap with any other known optical or radio continuum component. 

\begin{figure}[tb]
	\centering
	\includegraphics[clip=true,trim=4.cm 1.8cm 1cm 2.cm,scale=.35,angle=270]{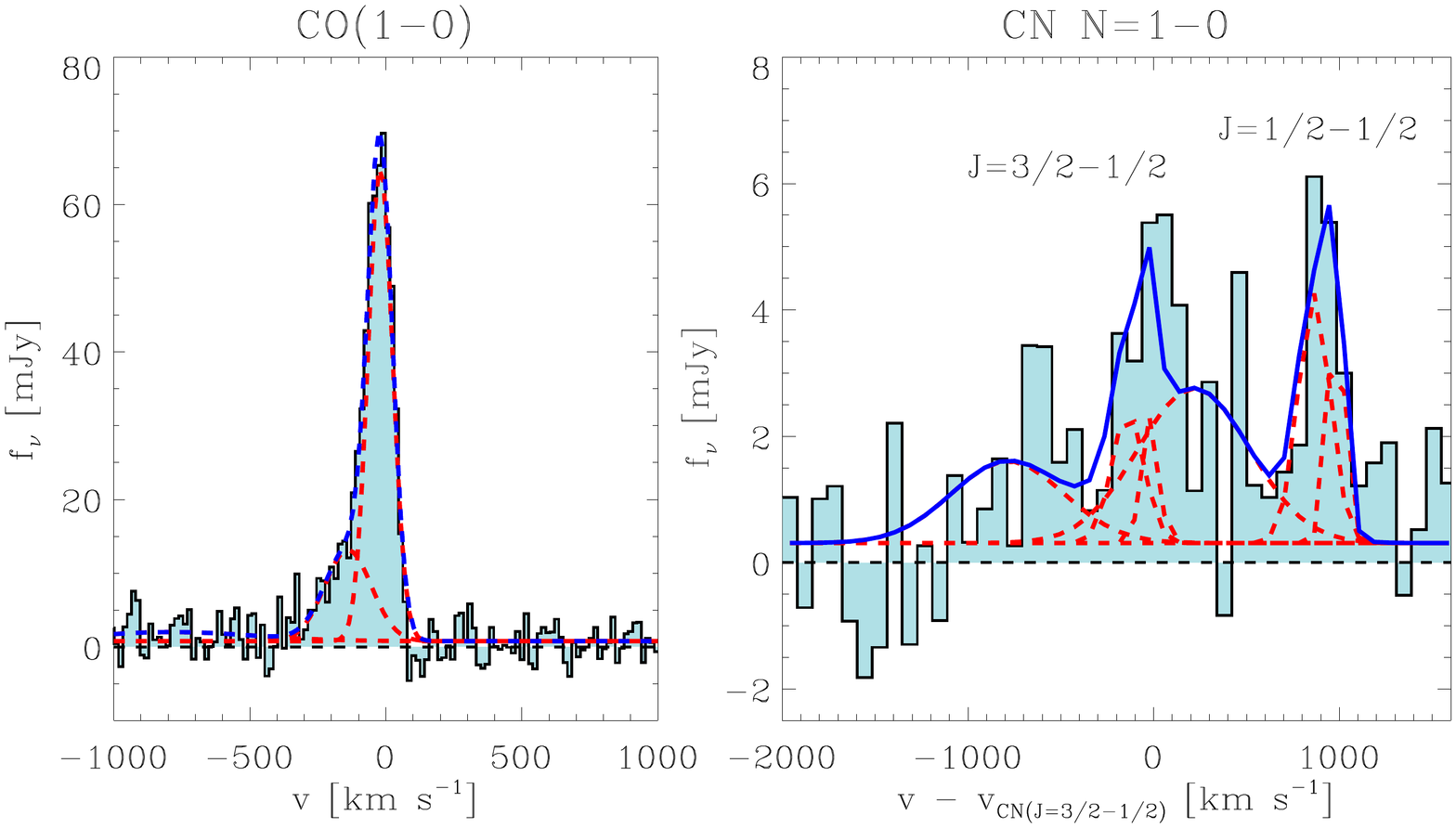}\quad
	\caption{CO(1-0) and CN(1-0) emission line spectra extracted from a circular aperture with 7$\arcsec$ radius centred at RA(J2000)=12:56:14.207, Dec(J2000)=56.52.30.994, sampling the northern feature of CN emission visible in the left panel of Fig.~\ref{fig:cn_maps}. 
	}\label{fig:spec_northern}
\end{figure}

To further study the origin of this emission, we show in Figure~\ref{fig:spec_northern} the CO(1-0) and CN(1-0) line spectra extracted from a $7\arcsec$-radius aperture, which samples the full extent of this northern CN component down to the $2\sigma$ level. The fluxes measured by integrating the spectra are $13.0\pm0.8$~Jy~\kms and $6\pm2$~Jy~\kms, respectively for CO(1-0) and CN(1-0). Most of the emission is blue-shifted in both tracers, with prominent broad blue-shifted components. The $L^{\prime}_{\rm CN(1-0)}/L^{\prime}_{\rm CO(1-0)}$ ratio is $0.47\pm0.16$, which is intermediate between the values measured in the narrow core of Mrk~231's integrated line emission and in the broad wings (see Fig.~\ref{fig:line_ratio_plots}). 

Hence, overall, the CN/CO line luminosity ratio and the broad and blueshifted line profiles suggest that the northern structure may trace an additional, extended ($r>5$~kpc) component of the molecular outflow. The nature of such emission needs to be investigated and confirmed with dedicated follow-up observations. However, the outflow hypothesis is also corroborated by the detection, both in molecular \citep{Aalto+12a,Cicone+12,Feruglio+15}, and in neutral atomic gas tracers \citep{Rupke+Veilleux11},
of high velocity gas up to $\sim2-3$~kpc North of the molecular disk, in a region bridging between the newly discovered CN-bright northern component and the well-known kpc-scale central molecular outflow.

 \section{Literature sample with CN(1-0) observations}\label{sec:AppendixB}
 

\longtab[1]{
\begin{longtable}{@{}lcccccccl@{}}
\caption{Literature Sample}
\label{table:literature}\\
\hline
\hline	
Name      	&	CO(1-0) 			&   CN(1-0)   &    HCN(1-0) 				&  HNC(1-0) 	&	HCO$^{+}$(1-0)  	& Type 	& Units	& Ref  \\
(1)			&	(2)					&    (3)			   		&   (4)						&	(5)			&	(6)			   		&	(7)     	 & (8)					& (9)           \\
\hline
\endfirsthead
\hline
\hline
Name      	&	CO(1-0) 				&   CN(1-0)   				&    HCN(1-0) 				&  HNC(1-0) 			&	HCO$^{+}$(1-0)  	 & Type & Units	& Ref  \\
(1)			&	(2)						&    (3)			   		&   (4)						&	(5)					&	(6)			   		  &	(7)     	 & (8)					& (9)           \\
\hline
\endhead
\hline
\multicolumn{9}{c}{\footnotesize Follows on next page}\\
\endfoot
\hline
\endlastfoot
\hline
Arp220          & $470.7\pm0.9$ & 44.3 $\pm$ 0.9  		&  54.9 $\pm$ 1.2 & 38.6 $\pm$ 1.7 & 22.9 $\pm$ 1.0   & I & Jy~km/s & a \\
IC694           & 283.5  		    & 13   $\pm$ 3 $^{(*)}$   &  11.8 $\pm$ 1.8 & 4.9  $\pm$ 1.2 & 19.2 $\pm$ 1.9   & I & Jy~km/s & b \\
NGC3690         & 290   $\pm$  30  & 61   $\pm$ 12 $^{(*)}$  &       14.       &   	$---$      &   11.1     	  & I & Jy~km/s & b,c \\
NGC1614         & 201.6 $\pm$  0.9 & 11.5 $\pm$ 1.1    	&  4.7  $\pm$ 1.7 & 1.4  $\pm$ 1.4 & 9.0  $\pm$ 1.8   & I & Jy~km/s & d \\
NGC2146         & 1150  $\pm$  30  & 49  $\pm$ 12 $^{(*)}$  &   $<18.7$ 	  & 7.5  $\pm$ 1.4 & 19.7 $\pm$ 3.3   & I & Jy~km/s & b,c \\
NGC2623         & 161				& 13.4		 			&  7.8	$\pm$ 0.7 &	 $---$ 	  	   & 4.0  $\pm$	0.7   & I & Jy~km/s & e,f \\
NGC6240         & 327  $\pm$   3   & 34.0 $\pm$ 1.7    	&  16.7 $\pm$ 0.9 &  3.3 $\pm$ 0.8 & 27.1 $\pm$ 1.0   & I & Jy~km/s & d,g \\
NGC3256$^{\rm N}$      & 1168 $\pm$ 	3 	& 280 $\pm$ 3   		&  197 $\pm$ 1.0  & 110.2 $\pm$ 0.9 & 205.3 $\pm$ 1.0 & N & K~km/s  &  h,i,j \\
NGC3256$^{\rm S}$      & 1344 $\pm$   3   & 73  $\pm$ 3   		&  57.0 $\pm$ 1.0 & 28.1  $\pm$ 0.9 & 87.9  $\pm$ 1.0 & N & K~km/s  &  h,i,j \\
NGC3256$^{\rm C}$      & 1037 $\pm$ 	2   & 95  $\pm$ 3  			&  93.2 $\pm$ 1.0 & 389.0 $\pm$ 0.9 & 139.5 $\pm$ 0.9 & OF & K~km/s & h,i,j  \\
NGC3256$^{\rm TNE}$    & 487 $\pm$    2   & 25  $\pm$ 3           &  37.3 $\pm$ 1.0 & 16.4  $\pm$ 0.9 & 71.9  $\pm$ 1.0 & TT & K~km/s &  h,i,j      \\
NGC3256$^{\rm TSE}$    & 749 $\pm$    2   & 25  $\pm$ 3           &  36.6 $\pm$ 1.0 & 13.8  $\pm$ 0.9 & 70.5  $\pm$ 1.0 & TT & K~km/s &  h,i,j     \\
NGC3256$^{\rm TSW}$    & 990 $\pm$    2   & 23  $\pm$ 3           &  51.9 $\pm$ 1.0 & 17.5  $\pm$ 0.9 & 84.5  $\pm$ 1.0 & TT & K~km/s &  h,i,j      \\
NGC3256$^{\rm OS}$     & 537 $\pm$    3   & 33  $\pm$ 3           &  40.4 $\pm$ 1.0 & 12.3  $\pm$ 0.9 & 64.6  $\pm$ 1.0 & OF & K~km/s &  h,i,j      \\
NGC3256$^{\rm SW}$     & 1317 $\pm$   2   & 67  $\pm$ 3           &  76.7 $\pm$ 1.0 & 33.3  $\pm$ 0.9 & 124.1 $\pm$ 1.0 & N  & K~km/s &  h,i,j       \\
NGC7130         & 257  $\pm$   14  & 20  $\pm$ 4 $^{(*)}$    &  16.5     		 & 10.0  $\pm$ 1.2 &  $---$     	 & I  & Jy~km/s &  b,c    \\
NGC1808         & 1900 $\pm$  140  & 154 $\pm$ 4 $^{(*)}$    &  107  $\pm$ 5   & 45    $\pm$ 11  &   59  $\pm$ 4   & I  & Jy~km/s &  b    \\
NGC660          & 659.0 $\pm$ 1.8  & 64.6 $\pm$ 1.6    	   &  28.7 $\pm$ 1.2 & 15.0  $\pm$ 1.2 &  29.9 $\pm$ 1.2 & I  & Jy~km/s & d   \\
NGC1068$^{\rm CND}$     & 155.0	$\pm$ 1.0  & 46.1 $\pm$ 0.5	   	   &  26.4 $\pm$ 0.6 &	5.6  $\pm$ 0.4 &  13.3 $\pm$ 0.7 & N  &	K~km/s	& k  \\
05414+5840  &  411.1     	   & 17.4 $\pm$ 0. $^{(*)}$ &  15 $\pm$ 3     &  $<3.8$    	   &  9    $\pm$ 3   & I  & Jy~km/s &   b       \\
12243-0036  &  165.4           & 14.1 $\pm$ 2.7 $^{(*)}$ &  13.0 $\pm$ 1.6 &  13.4 $\pm$ 1.6 &  9.7  $\pm$ 1.6 & I  & Jy~km/s &   b	       \\
M83             &  1121 $\pm$ 14   & 94.1 $\pm$ 0.7   	   &  50.5 $\pm$ 0.6 &  20.4 $\pm$ 0.4 &  46.1 $\pm$ 0.7 & I  & Jy~km/s &  a   \\
15107+0724  &  118.1  		   & 11.4 $^{(*)}$   	   &  13   $\pm$ 4   &  7    $\pm$ 4   &  7    $\pm$ 2   & I  & Jy~km/s &   b	   \\
17208-0014  &  171.            & 14.5     			   &  11.8 $\pm$ 0.8 &  9.1  $\pm$ 1.0 &  9.1  $\pm$ 0.8 & I  & Jy~km/s &   d,f   \\
NGC253$^{\rm tot}$  		&  870.	$\pm$   3  & 150.0 $\pm$ 1.4       &  70.0 $\pm$ 1.0 &  30.0 $\pm$ 1.0 &  59.0 $\pm$ 1.0 & I  &	K~km/s  &  k   \\
NGC253$^{\rm SB}$		&  16000 $\pm$ 700 & 2540 $\pm$ 80        &  2420 $\pm$ 110 &   $---$     	   &  2090 $\pm$ 100 & N  & K~km/s	&  l \\
NGC253$^{\rm r1}$		&  720 $\pm$ 70    & 33 $\pm$ 3 	       &  24   $\pm$ 2   & 	 $---$ 	  	   &  27   $\pm$ 3   & OF &	K~km/s	&  l \\
NGC253$^{\rm r10}$  	&  1920 $\pm$ 190  & 112 $\pm$ 9 	       &  179  $\pm$ 18  &	 $---$ 	  	   &  153  $\pm$ 15  & OF &	K~km/s  &  l  \\
NGC253$^{\rm r2}$		&  1920 $\pm$ 190. & 76 $\pm$ 6 	   	   &  119  $\pm$ 12  &	 $---$  	   &  76   $\pm$ 8   & OF & K~km/s  &  l  \\
IC342$^{\rm tot}$ 	& 160.0 $\pm$ 0.4  	   & 20.4 $\pm$ 0.4	   	   & 11.60 $\pm$ 0.10 &	4.5 $\pm$ 0.10 &  8.4  $\pm$ 0.10 & N &	K~km/s	&  k \\
M82         &  2658 $\pm$ 14.      & 213  $\pm$  2   	   & 117   $\pm$ 2    & 52.0 $\pm$ 1.1 &  167  $\pm$ 4    & I & Jy~km/s &  a   \\
NGC4945     &  9886.6  $\pm$ 28.4  & 1134 $\pm$ 30 $^{(*)}$  & 420   $\pm$ 8    &  225 $\pm$ 4   &  450  $\pm$ 4    & I & Jy~km/s &  b      \\
AM2055-425  &  49                 & $1.5 \pm0.3$ $^{(*)}$  &       $---$     &    $---$        &    $---$         & I & Jy~km/s &  m    \\
AM2246-490  &  34                 & 2.4 $\pm$ 0.3 $^{(*)}$ &       $---$     &    $---$        &    $---$         & I & Jy~km/s &  m    \\
VV114       &  613                & 6.0  ($^*$)  		   &       $---$     &    $---$        &    $---$         & I & Jy~km/s &  m \\
AM1300-233  &  89                 & 2.6 $\pm$ 0.3 $^{(*)}$ &       $---$     &    $---$    	   &    $---$  		  & I & Jy~km/s &  m \\
NGC1377     &  45 $\pm$ 5    	   & 0.48 $\pm$ 0.12 $^{(*)}$ &      $---$     &    $---$    	   &    $---$   	  & I & Jy~km/s &  m   \\
IC860       &  73.1   			   & 4.6      			   & 6.7 $\pm$ 0.7   &   3.6 $\pm$ 0.6 &  4.2  $\pm$ 0.6  & I & Jy~km/s &   d,f  \\
NGC3079     &  1090   $\pm$   50   & 32   $\pm$  4     	   & 25.0 $\pm$ 1.3  &   6.6 $\pm$ 0.9 & 28.1  $\pm$ 1.4  & I & Jy~km/s &   d    \\
NGC4194     &  171.3  $\pm$   1.0  & 10.0 $\pm$ 1.2       & 2.8 $\pm$ 0.4   &   1.5 $\pm$ 0.4 & 3.7   $\pm$ 0.4  & I & Jy~km/s &   d  \\
NGC7771     &  464.0  $\pm$   1.4  & 23 $\pm$  2   		   & 20.3 $\pm$ 0.9  & 12.0 $\pm$ 1.0  & 18.5  $\pm$ 0.9  & I & Jy~km/s &   d \\
NGC3556     &  251.2  $\pm$   1.8  & 6.0  $\pm$  0.8      & 3.2 $\pm$ 0.5   &   1.0 $\pm$ 0.5 & 5.1   $\pm$ 0.5  & I & Jy~km/s &   d \\
NGC7674     &  121.1  $\pm$   1.2  & 9.6  $\pm$  1.5      & $<1.1$       	 &   $<1.1$        & $<1.1$    		  & I & Jy~km/s &   d       \\
UGC2866     &  427.8  $\pm$   1.7  & 16.0 $\pm$  1.2      & 8.8 $\pm$   0.7 &   4.5 $\pm$ 0.8 & 12.9 $\pm$ 0.7   & I & Jy~km/s &   d    \\
UGC5101     &  96.9   $\pm$   1.8  & 22.1 $\pm$  1.6      & 6.6 $\pm$   0.9 &   5.4 $\pm$ 1.1 & 2.4  $\pm$ 1.0   & I & Jy~km/s &   d     \\
M51         &  249    $\pm$   8    & 38.0 $\pm$  0.4      & 26.7 $\pm$  0.5 &   8.9 $\pm$ 0.3 & 13.2 $\pm$ 0.3   & I & Jy~km/s &   a   \\
M51\_P1	    &  68.6	  $\pm$   0.3  & 3.2 $\pm$  0.08	   & 2.74 $\pm$ 0.09 & 0.96  $\pm$ 0.06	& 2.04 $\pm$ 0.08 &	I &	K~km/s	&   n   \\
M51\_P2    	&  54.1	  $\pm$   0.2  & 1. $\pm$ 0.3 	       & 1.70 $\pm$	0.10 & 0.63  $\pm$ 0.07	& 1.38 $\pm$ 0.09 &	I &	K~km/s	&   n	  \\
NGC7469     &  288    $\pm$    7   & 29 $\pm$ 2    	   	   & 11.7 $\pm$ 0.6  & 5.4   $\pm$ 0.6  & 14.3 $\pm$ 0.7  & I & Jy~km/s &   a    \\
IIIZw35		&  70.6   	           & 8.5 	   			   & 6.5  $\pm$	0.9	 &	 $---$	  	    & 3.0  $\pm$ 0.6  &	I &	Jy~km/s	&	e,f    \\
05189-2524  &  25.7	  	           & 6.7  	   			   & 4.4  $\pm$	0.8  &   2.3 $\pm$ 0.7	& 3.3  $\pm$ 0.8  & I & Jy~km/s	&   f,o     \\
09111-1007$^{\rm W}$	&  51	  	           & 6  				   & 3.6  $\pm$ 0.5  &    $<1.5$        & 5.0  $\pm$ 1.0  &	I &	Jy~km/s	&   f,o   \\
10173+0828	&  24.2   	           & 2   				   & 2.6  $\pm$	0.7	 &	  $---$ 	    & 1.7  $\pm$ 0.7  & I & Jy~km/s	&	e,f    \\
12224-0624  &  22.6	               & 3.1     			   & $<3.3$   		 &    $<4.0$        &  $<4.0$    	  & I & Jy~km/s &   f,o  \\
NGC4418		&  117	  	           & 15.1      			   & 12.9 $\pm$	0.9  & 6.0 $\pm$ 0.8    &  7.6 $\pm$ 0.8  & I & Jy~km/s &   d,f  \\
13120-5453	&  234	  	           & 39.2    	   		   &	$---$		 &	$---$	  		&  $---$   		  &	I & Jy~km/s	&	f	 \\
14378-3651	&  24.7	               & 3 				   &    $---$ 		 &	$---$ 	  		&  $---$    	  & I & Jy~km/s	&	f	 \\
20551-4250	&  54.9	   	           & 3 				   & 1.4 $\pm$ 0.1   &  0.37 $\pm$ 0.04	& 2.20 $\pm$ 0.10 &	I & Jy~km/s	&   f, p  \\
NGC7479		&  270	  	           & 9 $^{(*)}$	   	   &		$<20.2$  &	$---$	  		&   $---$	   	  &	I &	 Jy~km/s &	f, q                   \\
W51GMC		&  172	  $\pm$   4	   & 16.0 $\pm$  1.2	   & 5.5  $\pm$	0.2	 &	2.4  $\pm$  0.3	&   4.5	$\pm$ 0.2 &	C & K~km/s &  r	\\
W51GMCA		&  1617   $\pm$   11   & 124  $\pm$  7	       & 62.0 $\pm$	1.0	 &	40.  $\pm$  1.0	&   54.	$\pm$ 2	  &	C & K~km/s &	r	\\
W51GMCB		&  748	  $\pm$   6	   & 55   $\pm$  4	       & 30.  $\pm$	0.6	 &	16.3 $\pm$  0.6	&  23.4 $\pm$ 0.6 &	C & K~km/s &	r	\\
W51GMCC		&  369	  $\pm$   6    & 23   $\pm$  2	       & 11.8 $\pm$	0.5	 &	5.6  $\pm$  0.6	&  9.3	$\pm$ 0.5 &	C & K~km/s &  r	\\
W51GMCD		&  186	  $\pm$   6    & 16   $\pm$  2	       & 5.2  $\pm$	0.3	 &	2.2  $\pm$  0.3	&  4.5	$\pm$ 0.3 &	C & K~km/s &	r	\\
W51GMCE		&  77	  $\pm$   3	   & 13   $\pm$  1.4       & 2.1  $\pm$	0.3	 &	0.6  $\pm$  0.3	&  1.8	$\pm$ 0.2 &	C & K~km/s & r	\\
W51e1/e2	&  1375   $\pm$   6	   & 154  $\pm$  3	       & 77.2 $\pm$	0.5	 &	50.0  $\pm$  0.4	&  75.2	$\pm$ 0.5 &	C & K~km/s &	r	\\
OrionB		&  60.43  	 		   & 1.16 $^{(*)}$  		   &	1.54  	 	 &	0.445	        &   1.63	      &	C & K~km/s &	s \\
LMC\_N79	&  16.80  $\pm$ 0.06  & 0.68 $\pm$ 0.09 $^{(*)}$ & 1.51 $\pm$ 0.03 &  0.45 $\pm$ 0.03	& 2.797 $\pm$ 0.019 & C &	K~km/s & t   \\
LMC\_N44C	&  25.58  $\pm$ 0.05  & 0.51 $\pm$ 0.05 $^{(*)}$ & 1.192 $\pm$ 0.017 & 0.383 $\pm$ 0.013 & 2.164 $\pm$ 0.015 & C & K~km/s & t   \\
LMC\_N113	&  46.96  $\pm$ 0.07  & 0.90 $\pm$ 0.12 $^{(*)}$ & 3.15 $\pm$	0.05 &  1.39 $\pm$ 0.03	&  4.29	$\pm$ 0.04 & C & K~km/s &	 t  \\
OrionA	    & $3.0\pm1.2\times10^4$ & 140  $\pm$ 600 $^{(*)}$  & 1000 $\pm$	400	 &	$---$	        &  $---$	      &	C  & K~km/s~pc$^2$ &	u	 \\
\hline	
\end{longtable}
\tablefoot{Col.(1): source name. Cols.(2-6): integrated line fluxes of the corresponding transitions. Col.(7): source type. I= integrated observation; N= nucleus; OF=outflow region; TT: tidal tail; C = giant molecular cloud. Col.(8): units of the fluxes reported in Cols.(2-6); Col.(9): references: (a) \cite{Aladro+15}; (b) \cite{Baan+08}; (c) \cite{Aalto+02}; (d) \cite{Costagliola+11}; (e) \cite{Garcia-Burillo+12}; (f) Lutz et al. 2019 (in prep); (g) \cite{Papadopoulos+12_MNRAS}; (h) \cite{Harada+18}; (i) \cite{Sakamoto+14}; (j) Harada, Sakamoto (Private Comm.); (k) \cite{Takano+19}; (l) \cite{Meier+15}; (m) \cite{Wilson18}; (n) \cite{Watanabe+14}; (o) \cite{Privon+15}; (p) \cite{Imanishi+17}; (q) \cite{Gao+Solomon04}; (r) \cite{Watanabe+17}; (s) \cite{Pety+17}; (t) \cite{Nishimura+16}; (u) \cite{Kauffmann+17} and J. Kauffmann (priv. comm.); 
($^*$) In these cases only the flux of the CN(1-0) $J=3/2-1/2$ spingroup is available, and it has been multiplied by a factor of 1.5 to evaluate the total CN(1-0) flux.}
}

\end{appendix}

\end{document}